\documentclass[12pt]{article}
\usepackage{amssymb,amsmath,epsfig}
\usepackage{graphicx}
\usepackage{amsfonts,graphicx,amssymb,amsmath,amsthm,epsfig}
\usepackage{float}
\allowdisplaybreaks
\begin{document}
\title{\textbf{Charged Anisotropic Pulsar SAX J1748.9-2021 in Non-Riemannian Geometry}}
\author{M. Sharif$^{1,2}$ \thanks {msharif.math@pu.edu.pk}~ and Iqra Ibrar$^1$
\thanks{iqraibrar26@gmail.com}\\
$^1$ Department of Mathematics and Statistics, The University of Lahore,\\
1-KM Defence Road Lahore-54000, Pakistan.\\
$^2$ Research Center of Astrophysics and Cosmology, Khazar
University,\\ Baku, AZ1096, 41 Mehseti Street, Azerbaijan.}
\date{}
\maketitle

\begin{abstract}
In this paper, we investigate the impact of charge on the stability
of the pulsar star SAX J1748.9-2021 within the context of
$f(\mathbb{Q}, \mathcal{T})$ theory, where $\mathbb{Q}$ and
$\mathcal{T}$ represent the non-metricity scalar and the
energy-momentum tensor, respectively. To achieve this, we employ the
Krori-Barua metric ansatz with anisotropic fluid. We obtain exact
relativistic solutions for the corresponding field equations.
Additionally, we examine its physical and geometric properties using
astrophysical observations of the pulsar SAX J1748.9-2021. This
approach provides a basis for connecting several physical
quantities, including fluid parameters, anisotropy, mass-radius
relationship, compactness, redshift, \emph{energy}, the Zeldovich
and causality conditions, equation of state parameter, adiabatic
index and the Tolman-Oppenheimer-Volkoff equation. Our findings
align with observational evidence, indicating that the pulsar SAX
J1748.9-2021 remains both feasible and stable under this modified
theory.
\end{abstract}

\section{Introduction}

The splendor of the night sky is attributed to glowing, spherical
masses of plasma, bound by their own gravity. These entities are
essential for unraveling the structure and dynamics of the universe
\cite{1}. Stars serve as powerful energy sources, emitting a wide
range of radiation, such as light, heat, ultraviolet and X-rays.
These astronomical objects, comprising basically of helium and
hydrogen, are confined by gravitational forces that uphold their
structural integrity. Deep in their interiors, nuclear fusion
processes convert hydrogen into helium, unleashing immense energy
and giving rise to the brilliant luminosity that defines stars. By
studying stars, researchers gain valuable knowledge about their
formation and evolution. When a star depletes its nuclear energy,
the internal pressure declines, leading to an intensified
gravitational collapse. Eventually, this sequence culminates in the
star collapsing under its own gravity, leading to the formation of
dense remnants such as white dwarfs, neutron stars or black holes
\cite{2}.

White dwarfs are highly compact and dense, continuing to emit faint
light despite their diminutive size. Conversely, neutron stars
consist of an exceedingly dense aggregation of neutrons, signifying
an alternate phase in the lifecycle of stars. On the other hand,
black holes are regions of spacetime dominated by gravitational
forces so intense that even light cannot escape their grasp. In
contrast to ordinary stellar bodies, these compact entities are
devoid of the requisite thermodynamic pressure to counteract
gravitational implosion. Rather, their immense density is governed
by quantum mechanical principles, particularly the Pauli exclusion
principle, which prevents identical fermions from existing in the
same quantum state. This principle generates a form of degeneracy
pressure that halts further collapse \cite{3}. Consequently, these
remnants are extraordinarily compact yet exhibit immensely intense
gravitational and magnetic fields.

Scholars suggest that compact stellar remnants emerge from the
aftermath of supernova cataclysms \cite{4}. This proposition finds
additional corroboration in the detection of neutron stars that
rotate at remarkable velocities \cite{5}. These dense celestial
bodies are highly intriguing as they provide insights into the
behavior of matter in extreme environments, rendering them a vital
focus of study in both astronomy and physics. Pulsars are
distinguished by their periodic emission of intense bursts of
radiation. A notable example, Pulsar SAX J1748.9-2021 (PS) is a
fascinating neutron star. It belongs to a category of millisecond
pulsars, characterized by its rapid rotation and emission of X-ray
pulses. This PS exhibits complex behavior, including episodes of
X-ray outbursts resulting from the accretion of matter from a
companion star in a binary system. Such interactions make PS an
important object of study for understanding the physics of
accretion, stellar evolution and the extreme conditions around
neutron stars. Additionally, observations of this PS help to test
theories of gravity in strong-field regimes, offering insights into
the nature of dense matter and spacetime. Ongoing research on
pulsars remains exceptionally active, with every new finding
broadening our knowledge of these intriguing cosmic phenomena
\cite{6}.

General Relativity (GR) has been pivotal in elucidating our
comprehension of gravity, characterizing it as the distortion of
spacetime geometry instigated by the distribution of mass and energy
\cite{7}. Despite its success in explaining a wide range of
astrophysical phenomena, GR faces challenges at cosmological scales,
particularly in accounting for the observed acceleration of the
universe's expansion without invoking dark energy (DE) or
modifications in the matter content. This has motivated the
development of alternative theories of gravity, one of which is the
$f(\mathbb{Q}, \mathcal{T})$ theory. This theory extends the
framework of GR by introducing non-metricity, a geometric quantity
that defines how spacetime is altered without relying on curvature
or torsion. Unlike traditional GR, which focuses on the curvature of
spacetime, this theory employs non-metricity to describe
gravitational interactions. By incorporating the trace of the
energy-momentum tensor (EMT), the $f(\mathbb{Q}, \mathcal{T})$
theory allows a coupling between matter and geometry, thereby
modifying the gravitational field equations \cite{9}. This coupling
provides new dynamics that can potentially address unexplained
cosmological phenomena, such as the universe accelerated expansion,
without the need for additional exotic matter components. Thus,
$f(\mathbb{Q}, \mathcal{T})$ offers a versatile and intriguing
extension to GR, opening up new possibilities for exploring the
complexities of cosmic evolution.

The investigation of this theory is motivated by its significant
theoretical impact and its crucial contribution to explaining
various astronomical phenomena. Arora et al. \cite{10} explored
cosmic acceleration, revealing its potential occurrence without DE,
while Bhattacharjee \cite{11} investigated the effects of this
gravity, concluding that it notably influences tidal forces. Godani
and Samanta \cite{12} explored cosmic evolution by employing various
astrophysical parameters and concluded that this framework
effectively accounts for the universe persistent accelerated
expansion. Agrawal et al. \cite{13} analyzed the dynamical
properties within this theoretical model, offering insights into its
influence on the universe evolution and the processes occurring in
the early stages of cosmic history. Shiravand et al. \cite{14}
investigated the implications of the theory for cosmic inflation.
Tayde et al. \cite{15} explored the potential for wormhole
configurations, assessing its viability in supporting such exotic
structures in the cosmos. Pradhan et al. \cite{16} assessed the
stability and feasibility of \textit{gravastars} under this
alternative gravitational framework, examining its capacity to
support such exotic astrophysical objects. In our latest study, we
have explored the reformulation of the ghost DE model using this
framework \cite{17}.

The unique characteristics and complex structures of PS have
garnered considerable attention, driving significant progress in
alternative theoretical frameworks. Mak and Harko \cite{7-c}
assessed the equilibrium of pulsars by assessing their energy
thresholds and equilibrium states, utilizing the speed of sound as a
pivotal parameter. Kramer et al. \cite{7-d} conducted an in-depth
study of the pulsar, emphasizing its extraordinary potential for
performing rigorous evaluations of GR and alternative gravitational
theories. Owing to its distinct features and relative proximity,
this system offers the possibility of more accurate assessments than
current solar system experiments, potentially challenging prevailing
assumptions about pulsar formation. G$\ddot{u}$ver and $\ddot{O}$zel
\cite{34} applied temporal-resolution spectroscopy to investigate
X-ray bursts emanating from PS. Mafa Takisa et al. \cite{34-a}
investigated the \textit{Einstein-Maxwell system} for anisotropic,
charged dense stellar bodies, demonstrating that the presence of
charge leads to an increase in mass, with results that align well
with observed data for various compact stars. Maurya and Gupta
\cite{7-ab} extended their model to derive solutions for charged
fluid distributions, revealing that both anisotropy and electric
field intensity escalate. This insight highlights the intricate
nature of these stellar entities. Furthermore, recent research has
examined the impacts of $f(\mathbb{R})$ and $f(\mathbb{Q},
\mathcal{T})$ gravitational frameworks on PS \cite{7-e}.

charged dense stellar objects are exotic astrophysical objects
characterized by their extremely high density and significant net
electric charge. These stars, often hypothesized to form under
intense astrophysical conditions, exhibit a delicate balance between
gravitational collapse and electrostatic repulsion. The presence of
charge alters the internal structure and stability of these stars,
as the electrostatic forces provide an additional counterbalance to
gravity, potentially allowing them to attain more compact
configurations than their neutral counterparts, such as PS
\cite{11f}. Theoretical models suggest that the charge in such stars
could arise from the separation of charges in the dense matter or
from the accumulation of charged particles on the star's surface
\cite{12f}. This electric charge also affects the star's
electromagnetic fields and the surrounding spacetime, influencing
the dynamics of nearby matter and radiation and potentially leading
to unique observational signatures that could distinguish charged
compact stars from other compact objects in the universe \cite{13f}.

This paper investigates the influence of charge on the stability and
feasible aspects of anisotropic PS within the framework of
$f(\mathbb{Q},\mathcal{T})$ gravity. The paper is structured as
follows. Section \textbf{2} outlines the fundamental principles of
$f(\mathbb{Q},\mathcal{T})$ gravity and its associated field
equations in the presence of charge. Additionally, it details the
formulation and solution of the Einstein-Maxwell equations within
the context of this theory. In Section \textbf{3}, we derive the
field equations for a specific $f(\mathbb{Q},\mathcal{T})$ model and
employ the Krori-Barua (KB) ansatz. The matching conditions are then
utilized to identify the unknown constants within the KB ansatz.
Sections \textbf{4} and \textbf{5} employ observational data of
charged PS to derive the key physical parameters essential for the
analysis. The model's stability is then evaluated in the light of
these physical constraints. The paper concludes by summarizing our
results.

\section{The $f(\mathbb{Q},\mathcal{T})$ Formalism}

Under the paradigm of $f(\mathbb{Q},\mathcal{T})$ gravity, the
modified action, which seamlessly incorporates the electromagnetic
field Lagrangian $\mathbb{L}_{e}$ and the matter Lagrangian
$\mathbb{L}_{m}$ is conveyed by the following expression \cite{9}
\begin{equation}\label{22}
S=\frac{1}{2\kappa}\int f(\mathbb{Q},
\mathcal{T})\sqrt{-g}d^{4}x+\int
(\mathbb{L}_{m}+\mathbb{L}_{e})\sqrt{-g}d^{4}x.
\end{equation}
In this context, $g$ represents the determinant of the metric
tensor, $\kappa$ denotes the coupling constant (set to unity) and
the Lagrangian for the electromagnetic field is expressed as
\begin{equation}
\mathbb{L}_{e} =\frac{-1}{16\pi} F^{\varsigma\nu}F_{\varsigma\nu}.
\end{equation}
In this expression, $F^{\varsigma\nu} = \psi_{\varsigma,\nu} -
\psi_{\nu,\varsigma}$ represents the Maxwell field tensor, with
$\psi_{\varsigma}$ being the four-potential. Furthermore,
\begin{equation}\label{23}
\mathbb{Q}=-g^{\xi\varrho}(\mathrm{L}^{\rho}_{\mu\xi}\mathrm{L}^{\mu}_{\varrho\rho}
-\mathrm{L}^{\rho}_{\mu\rho}\mathrm{L}^{\mu}_{\xi\varrho}),
\end{equation}
where
\begin{equation}\label{24}
\mathrm{L}^{\rho}_{\mu\varpi}=-\frac{1}{2}g^{\rho\lambda}
(\nabla_{\varpi}g_{\mu\lambda}+\nabla_{\mu}g_{\lambda\varpi}
-\nabla_{\lambda}g_{\mu\varpi}).
\end{equation}
The following expressions define the traces associated with the
non-metricity tensor
\begin{align}\label{25}
\mathbb{Q}_{\rho}= \mathbb{Q}^{~\xi}_{\rho~\xi}, \quad
\tilde{\mathbb{Q}}_{\rho}= \mathbb{Q}^{\xi}_{\rho\xi}.
\end{align}
The superpotential related to $\mathbb{Q}$ is expressed as follows
\begin{align}\label{26}
P^{\rho}_{\xi\varrho}&=-\frac{1}{2}\mathrm{L}^{\rho}_{\xi\varrho}
+\frac{1}{4}(\mathbb{Q}^{\rho}-\tilde{\mathbb{Q}}^{\rho})
g_{\xi\varrho}- \frac{1}{4} \delta
^{\rho}\;_{({\xi}}\mathbb{Q}_{\varrho)}.
\end{align}
Moreover, the description of $\mathbb{Q}$, as formulated based on
the superpotential, can be expressed in the following form \cite{17}
\begin{align}\label{27}
\mathbb{Q}=-\mathbb{Q}_{\rho\xi\varrho}P^{\rho\xi\varrho}
=-\frac{1}{4}(-\mathbb{Q}^{\rho\xi\varrho}\mathbb{Q}_{\rho\varrho\xi}
+2\mathbb{Q}^{\rho\varrho\xi}\mathbb{Q}_{\xi\rho\varrho}
-2\mathbb{Q}^{\varrho}\tilde{\mathbb{Q}}_{\varrho}
+\mathbb{Q}^{\varrho}\mathbb{Q}_{\varrho}).
\end{align}

The gravitational equations are derived by taking the variation of
the action $S$ concerning the metric tensor and equating the
resulting formulation to zero
\begin{align}\nonumber
\delta S&=0=\int \frac{1}{2}\delta
[f(\mathbb{Q},\mathcal{T})\sqrt{-g}]+\delta[\mathbb{L}_{m}\sqrt{-g}]d^{4}x
\\\nonumber 0&=\int \frac{1}{2}\bigg( \frac{-1}{2} f g_{\xi\varrho}
\sqrt{-g} \delta g^{\xi\varrho} + f_{\mathbb{Q}} \sqrt{-g} \delta
\mathbb{Q} + f_{\mathcal{T}} \sqrt{-g} \delta
\mathcal{T}\bigg)\\\label{28}&-\frac{1}{2} \mathcal{T}_{\xi\varrho}
\sqrt{-g} \delta g^{\xi\varrho}d^ {4}x.
\end{align}
In addition, we define
\begin{align}\label{29}
\mathcal{T}_{\xi\varrho} &= \frac{-2}{\sqrt{-g}} \frac{\delta
(\sqrt{-g} \mathbb{L}_{m})}{\delta g^{\xi\varrho}}, \quad
\Theta_{\xi\varrho}= g^{\rho\mu} \frac{\delta
\mathcal{T}_{\rho\mu}}{\delta g^{\xi\varrho}}.
\end{align}
This leads to $ \delta \mathcal{T} = \delta
(\mathcal{T}_{\xi\varrho}g^{\xi\varrho}) = (\mathcal{T}_{\xi\varrho}
+ \Theta_{\xi\varrho})\delta g^{\xi\varrho}$. Including these
factors into Eq.\eqref{28} results in the following expression
\begin{eqnarray}\nonumber
\delta S =0&=&\int \frac{1}{2}\bigg\{\frac{-1}{2}f
g_{\xi\varrho}\sqrt{-g} \delta g^{\xi\varrho} +
f_{\mathcal{T}}(\mathcal{T}_{\xi\varrho}+
\Theta_{\xi\varrho})\sqrt{-g} \delta g^{\xi\varrho}
\\\nonumber
&-&f_{\mathbb{Q}} \sqrt{-g} (P_{\xi\rho\mu}
\mathbb{Q}_{\varrho}~^{\rho\mu}- 2\mathbb{Q}^{\rho\varrho}~_{\xi}
P_{\rho\mu\varrho}) \delta g^{\xi\varrho}+2f_{\mathbb{Q}} \sqrt{-g}
P_{\rho\xi\varrho} \nabla^{\rho} \delta g^{\xi\varrho}
\\\label{30}
&+&2f_{\mathbb{Q}}\sqrt{-g}P_{\rho\xi\varrho} \nabla^{\rho} \delta
g^{\xi\varrho} \bigg\}- \frac{1}{2}
\mathcal{T}_{\xi\varrho}\sqrt{-g} \delta g^{\xi\varrho}d^ {4}x.
\end{eqnarray}
Integrating the term $2f_{\mathbb{Q}} \sqrt{-g} P_{\rho\xi\varrho}
\nabla^{\rho} \delta g^{\xi\varrho}$ and applying the boundary
conditions yield the expression $-2 \nabla^{\rho} (f_{\mathbb{Q}}
\sqrt{-g} P_{\rho\xi\varrho}) \delta g^{\xi\varrho}$. Here, the
terms $f_{\mathbb{Q}}$ and $f_{\mathcal{T}}$ indicate the partial
derivatives of the function concerning $\mathbb{Q}$ and
$\mathcal{T}$, respectively. Therefore, the field equations can be
written as follows
\begin{eqnarray}\nonumber
\mathcal{T}_{\xi\varrho}+E_{\xi\varrho}&=&\frac{-2}{\sqrt{-g}}
\nabla_{\rho} (f_{\mathbb{Q}}\sqrt{-g} P^{\rho}_{\xi\varrho})-
\frac{1}{2} f g_{\xi\varrho} + f_{\mathcal{T}}
(\mathcal{T}_{\xi\varrho} + \Theta_{\xi\varrho})
\\\label{31}
&-&f_{\mathbb{Q}} (P_{\xi\rho\mu} \mathbb{Q}_{\varrho}~^{\rho\mu}
-2\mathbb{Q}^{\rho\mu}~_{\xi} P_{\rho\mu\varrho}).
\end{eqnarray}

The reason for using non-Riemannian geometry, especially
$f(\mathbb{Q},\mathcal{T})$ gravity, can be explained as follows.
General Relativity is built upon the Levi-Civita connection, which
is both torsion-free and metric-compatible. In contrast,
non-Riemannian geometries relax these constraints, allowing for the
incorporation of additional geometric structures such as torsion and
non-metricity. These are mathematically represented by tensors like
the contortion tensor and the disformation tensor, which describe
deviations from the symmetric connection used in GR. From a physical
perspective, this extended geometric framework enables alternative
gravitational models such as $f(\mathbb{Q},\mathcal{T})$ gravity, to
investigate fundamental questions in cosmology. It provides a means
to study dark energy (DE) and the accelerated expansion of the
universe by introducing additional degrees of freedom. This
flexibility allows for the development of theories that could align
more closely with observational data and offer a deeper
understanding of gravity on cosmological scales. Thus, incorporating
non-Riemannian geometry into $f(\mathbb{Q},\mathcal{T})$ gravity
broadens the theoretical foundation of gravitational physics,
addressing some of the unresolved challenges in our current
understanding.

To assess the equilibrium properties of a stellar core, we adopt a
\textit{spherically symmetric} model. The corresponding line element
for this configuration can be expressed as
\begin{equation}\label{a1}
ds^{2}=-e^{\alpha(r)}dt^{2}+e^{\beta(r)}dr^{2}
+r^{2}(d\theta^{2}+\sin^{2}d\phi^{2}).
\end{equation}
The stellar matter distribution is considered anisotropic and is
defined by the following properties
\begin{equation}\label{a2}
{\mathcal{T}}_{\xi\varrho}=(\rho+p_{t})u_{\xi}u_{\varrho}+p_{t}g_{\xi\varrho}-\sigma
k_{\xi}k_{\varrho}.
\end{equation}
In this framework, $u_{\xi}$ and $u_{\varrho}$ refer to the
four-velocity components of the fluid and they follow the
normalization condition $u^\xi u_\xi = -1$. Similarly, $k_{\xi}$ is
identified as a unit radial four-vector, satisfying $k^{\xi} k_{\xi}
= 1$. The quantities of interest include the energy density $\rho$,
the radial pressure $p_{r}$ and the tangential pressure $p_{t}$. The
anisotropy factor, $\chi$, is specified as $\chi = p_{t} - p_{r}$.
Using the given metric, the four-velocity $u^{\xi}$ can be expressed
as $e^{-\alpha} \delta^\xi_{t}$ and the radial vector $k^\xi$ can be
written as $e^{-\beta} \delta^\xi_{r}$. As a result, the
Eq.\eqref{a2} can be outlined as
\begin{equation}\label{a3}
\mathcal{T}=3p_{r}-\rho+2\sigma.
\end{equation}
The electromagnetic field's stress-energy tensor is formulated as
\begin{equation}
E_{\xi\varrho} = \frac{1}{4\pi} \big( F^{\nu}_{\xi} F_{\nu\varrho} -
\frac{1}{4} g_{\xi\varrho} F_{\varsigma\nu} F^{\varsigma\nu} \big).
\end{equation}
The Maxwell equations are formulated as
\begin{equation}
(\sqrt{-g} F_{\xi\varrho})_{;\varrho} = 4\pi J_\xi \sqrt{-g}
F_{[\xi\varrho;\delta]} = 0.
\end{equation}
Within these equations, the electric four-current is represented as
$J_\xi = \sigma u_\xi$, where $\sigma$ stands for the charge
density.

The strength of the electric field is given by
\begin{equation}
E(r) = \frac{e^{\frac{a +b}{2}}}{r^2} q(r).
\end{equation}
Here, $q(r)$ represents the total charge enclosed within a sphere of
radius $r$ and is defined as
\begin{equation}
q(r) = 4\pi \int_0^r \sigma r^2 e^b dr,
\end{equation}
which implies that
\begin{equation}
\sigma = \frac{e^{-b}}{4\pi r^2} \frac{dq(r)}{dr}.
\end{equation}
For this study, we employ a specific expression for the charge,
defined as \cite{7-ee}
\begin{equation}\label{11-aa}
q = q_{0} r^{3},
\end{equation}
In this context, $q_{0}$ represents the charge intensity. A value of
$q_{0} = 0$ signifies an uncharged situation. The field equations
\eqref{31} yield the following non-zero components
\begin{align}\nonumber
\rho&= \frac{1}{2r^{2}e^{\beta}} \Bigg[ f_{\mathbb{Q}\mathbb{Q}}2r
\mathbb{Q}'(r) (e^{\beta} - 1) + f_\mathbb{Q} \Big( (e^{\beta} - 1)
(2 + r\alpha '(r)) + (e^{\beta} + 1) r \beta '(r) \Big)
\\\label{1-a}
&+ f r^2 e^{\beta} \Bigg] - \frac{1}{3} f_\mathcal{T} (3\rho+ p_r +
2p_t)- \frac{q^2}{r^4},
\\\nonumber
p_r &= -\frac{1}{2r^{2}e^{\beta}}\Bigg[2r \mathbb{Q}'(r)
f_{\mathbb{Q}\mathbb{Q}} (e^{\beta} - 1) + f_\mathbb{Q} \Big(
(e^{\beta} - 1) (2 + r\alpha '(r)) + r b') - 2r \alpha '(r)) \Big)
\\\label{1-b}
&+ f r^2 e^{\beta} \Bigg] + \frac{q^2}{r^4} + \frac{2}{3}
f_\mathcal{T} (p_t - p_r),
\\\nonumber
p_t &= -\frac{1}{4r e^{\beta}} \Bigg[ -2 r \mathbb{Q}'(r) \alpha'
f_{\mathbb{Q}\mathbb{Q}} + f_\mathbb{Q} \Big( 2\alpha' (e^{\alpha} -
2) - r \beta'^2 +b' (2 e^{\alpha} + r \alpha') - 2r \alpha' \Big)
\\\label{1-c}
&+ 2 f r e^{\beta} \Bigg] - \frac{q^2}{r^4} + \frac{1}{3}
f_\mathcal{T} (p_r - p_t).
\end{align}
Here, the prime symbol signifies differentiation concerning the
radial coordinate $r$.

\section{The $f(\mathbb{Q}, \mathcal{T})$ Gravity Paradigm}

Next, we explore the impact of $f(\mathbb{Q}, \mathcal{T})$ on the
geometric structure of a charged PS. In this analysis, we utilize a
specific form of $f(\mathbb{Q}, \mathcal{T})$, as outlined in
\cite{10-b}
\begin{align}\label{A}
f(\mathbb{Q}, \mathcal{T})&=A\mathbb{Q} + B\mathcal{T}.
\end{align}
In this context, $A$ and $B$ are arbitrary constants, where
$f_\mathbb{Q} = A$, $f_{\mathbb{QQ}} = 0$, $f_\mathcal{T} = B$ and
$f_{\mathcal{TT}} = 0$. This particular cosmic framework has been
thoroughly explored in numerous studies \cite{10-c}. Inserting these
values into Eqs. \eqref{1-a}-\eqref{1-c} yields the ensuing outcomes
\begin{align}\nonumber
\rho&=\bigg[e^{-\beta (r)} \big(\big(\alpha '(r)B \big(\big(e^{\beta
(r)}-1\big) \big(2 r^2-1+ e^{\beta (r)}\big)+r \beta
'(r)\big)+\big(B \big(e^{\beta (r)}
\\\nonumber
&\times\big(2 r^2\big(e^{5+\beta (r)}\big)-1\big)\big)\big) \beta
'(r)-2 B r \alpha ''(r)+B(-r) \alpha '(r)^2A  r^3 \big)
\\\nonumber
&-2 \big(B \big(6 r^2 e^{\beta (r)}-7\big)-6\big) \big(e^{\beta (r)}
\big(q^2-A r^2\big)+A r^2\big)\big)\bigg]\bigg[2 r^4 \big(B \big(-8
B
\\\label{a-1}
&+2 (4 B +3) r^2 e^{\beta (r)}-15\big)-6\big)\bigg]^{-1},
\\\nonumber
p_{r}&= \bigg[e^{-\beta (r)} \big(4 B  r^2 e^{2 \beta (r)} \big((7 B
+3) q^2+A  (B +3) r^2\big)-2 e^{\beta (r)} \big((B (3 B +13)+6) q^2
\\\nonumber
&+A  r^2 \big(B \big(13 B+2 (B +3) r^2+17\big)+6\big)\big)+2 A (B
(13 B +17)+6) r^2+A r^3
\\\nonumber
&\times\big(-2B  r \alpha ''(r)\big(3 B +2 (4 B +3) e^{\beta (r)}r^2
+2\big)+\alpha '(r) \big(B (27 B +47)+B r \big(3 B
\\\nonumber
&+2 (4 B +3)r^2 e^{\beta (r)}+2\big) \beta '(r)+2 (B +1) (11 B +6)
r^2 e^{2 \beta (r)}-e^{\beta (r)} \big((B +1)
\\\nonumber
&\times (11 B +6)+2 (B (19 B +23)+6) r^2\big)+18\big)+\alpha '(r)^2
B (-r) \big(3 B +2 (4 B +3)
\\\nonumber
&\times r^2 e^{\beta (r)}+2\big)+\big(B  (21+13 B)+2 (B +1) (11 B
+6) r^2 e^{2 \beta (r)}-e^{\beta (r)} \big((B +1)
\\\nonumber
&\times (11B +6)+2 (B +2) (B +3) r^2\big)+6\big) \beta
'(r)\big)\big)\bigg]\bigg[2 (B +1) r^4 \big(B  \big(-8 B
\\\label{a-2}
&+2 (4 B +3)r^2 e^{\beta (r)}\big)\big)\bigg]^{-1},
\\\nonumber
p_{t}&=\bigg[e^{-\beta (r)} \big(8 B  r^2 e^{2 \beta (r)} \big(A B
r^2-(5 B +3) q^2\big)+4 e^{\beta (r)} \big((B (9 B +17)+6) q^2-A B
r^2
\\\nonumber
&\big(-2 \big(B +2 B r^2+2\big)\big) +4 A  B (B +2) r^2+A r^3 \big(2
\big(r \alpha ''(r) \big(B \big(-6 B +2 (4 B +3)
\\\nonumber
&r^2 e^{\beta (r)}\big)\big)+\big(B \big(B +e^{\beta (r)} \big(B -2
r^2 \big(B +(B +1) e^{\beta (r)}\big)\big)\big)\big)
\\\nonumber
&\times\beta '(r)\big)+\alpha '(r) \big(r \big(B \big(6 B -2 (4 B
+3) r^2 e^{\beta (r)}+13\big)+6\big) \beta '(r)+2 B \big(-9 B
\\\nonumber
&+e^{\beta (r)} \big(B -2 r^2 \big(-5 B +(B +1) e^{\beta
(r)}-4\big)+1\big)-16\big)-12\big)+r \alpha '(r)^2 \big(B
\\\nonumber
&\times\big(-6 B +2 (4 B +3) r^2 e^{\beta
(r)}-13\big)-6\big)\big)\big)\bigg]\bigg[4 (B +1) r^4 \big(B \big(-8
B +2
\\\label{a-3}
&\times (4 B +3)r^2 e^{\beta (r)}\big)\big)\bigg].
\end{align}
In order to scrutinize stellar evolution, a series of well-founded
assumptions are imposed on the metric potentials $\alpha$ and
$\beta$. By leveraging the KB spacetime, an exhaustive analysis of
stellar configurations is conducted within the ambit of modified
gravitational theory.

\subsection{The KB Metric Potentials}

The KB metric potentials are essential functions characterizing the
geometry of spherically symmetric spacetimes. These potentials
dictate the gravitational field within the stellar interior and
influence key properties such as $p$, $\rho$ and $\chi$. In modified
gravity theories, the KB metric potentials serve as vital components
for constructing stellar models, providing insights into how
deviations from GR affect the internal structure and evolution of
compact objects. The KB spacetime's metric potentials are
comprehensively expressed in the following manner \cite{10-d}
\begin{equation}\label{4-a}
\alpha(r)=\gamma_{0} r^2+\gamma_{1},\quad \beta(r)=\gamma_{2}r^2.
\end{equation}
Here $\gamma_{0}, \gamma_{1},$ and $\gamma_{2}$ are unknown
constants to be determined through the application of junction
conditions. This solution guarantees that the star's interior
remains smooth and devoid of singularities.

\subsection{Matching Conditions}

These conditions are essential in the analysis of stellar
configurations within GR, as they ensure a smooth transition between
the star's internal solution and the surrounding external vacuum
solution. In this context, vacuum solutions from non-Riemannian
geometry are considered equivalent to those from $f(\mathbb{Q},
\mathcal{T})$ theory. Therefore, the model described in \eqref{A}
applies, with the exterior spacetime characterized by the vacuum
charged Reissner-Nordstr$\ddot{o}$m solution. Thus, the exterior
spacetime is modeled using the following expression
\begin{equation}\label{c-1}
ds^{2}=-\bigg(1+\frac{\mathbf{Q}^{2}}{r^{2}}-\frac{2\mathbf{M}}{r}\bigg)dt^{2}
+\bigg(1+\frac{\mathbf{Q}^{2}}{r^{2}}-\frac{2\mathbf{M}}{r}\bigg)^{-1}dr^{2}
+r^{2}(d\theta^{2}+\sin^{2}\theta d\phi^{2}).
\end{equation}
In this context, $\mathbf{M}$ denotes the mass of the star. Imposing
the seamless continuity of the metric functions from Eqs.\eqref{a1}
and \eqref{c-1} at the outer boundary $(r = \mathbf{R})$ results in
\begin{eqnarray}\label{cc-1}
g_{tt}&=&e^{\gamma_{0}\mathbf{R}^2
+\gamma_{1}}=\bigg(1+\frac{\mathbf{Q}^{2}}{\mathbf{R}^{2}}-\frac{2\mathbf{M}}{\mathbf{R}}\bigg),
\\
\label{1b}
g_{rr}&=&e^{\gamma_{2}\mathbf{R}^2}=
\bigg(1+\frac{\mathbf{Q}^{2}}{\mathbf{R}^{2}}-\frac{2\mathbf{M}}{\mathbf{R}}\bigg)^{-1},
\\
\label{1c}
g_{tt,r}&=&2\gamma_{0}\mathbf{R} e^{\gamma_{0}\mathbf{R}^2
+\gamma_{1}}=\frac{2\mathbf{M}}{\mathbf{R}^2}-\frac{2\mathbf{Q}^2}{\mathbf{R}^3}.
\end{eqnarray}
Solving \eqref{cc-1}-\eqref{1c} and applying the condition
$p_{r}(r=\mathbf{R})=0$, the constants $\gamma_{0}$, $\gamma_{1}$
and $\gamma_{2}$ can be found as follows
\begin{eqnarray}\label{2a}
\gamma_{0}&=&
\bigg(-\frac{2\mathbf{Q}^2}{\mathbf{R}^4}+\frac{2\mathbf{M}}{\mathbf{R}^3}\bigg)\bigg(-\frac{2\mathbf{M}}{\mathbf{R}}+1
+\frac{\mathbf{Q}^{2}}{\mathbf{R}^{2}}\bigg)^{-1},\\
\label{2b}
\gamma_{1}&=&\ln\bigg(1-\frac{2\mathbf{M}}{\mathbf{R}}+\frac{\mathbf{Q}^{2}}{\mathbf{R}^{2}}\bigg)
-\bigg(\frac{2\mathbf{M}}{\mathbf{R}^3}-\frac{2\mathbf{Q}^2}{\mathbf{R}^4}\bigg)\bigg(1-\frac{2\mathbf{M}}{R}
+\frac{\mathbf{Q}^{2}}{\mathbf{R}^{2}}\bigg)^{-1},\\
\label{2c} \gamma_{2}&=&\frac{\ln \bigg(\frac{\mathbf{R}^2}{-2
\mathbf{M}
\mathbf{R}+\mathbf{Q}^2+\mathbf{R}^2}\bigg)}{\mathbf{R}^2}.
\end{eqnarray}

\section{Stability Features of SAX J1748.9-2021}

Within this analysis, we delve into the observational data
pertaining to the mass and radius of the PS under the ambit of
$f(\mathbb{Q}, \mathcal{T})$ gravity. The selection of this
particular PS is justified by the availability of extensive
spectroscopic observations, providing precise measurements during
its X-ray burst phases. Specifically, the mass is estimated to be
$\mathbf{M} = 1.81 \pm 0.3 \, \mathbf{M}_\odot$, while the radius is
determined as $\mathbf{R} = 11.7 \pm 1.7 \, \text{km}$ \cite{10-g}.
By leveraging these parameters, we can rigorously explore the
intricate physical attributes of the star and assess its stability.

\subsection{Material Constituent}

In the examination of stellar configurations, fundamental quantities
$\rho$, $p_{r}$, $p_{t}$ and $\chi$ are paramount in comprehending
the star's dynamics and equilibrium stability. Gaining profound
insight into the star's internal architecture and the dynamics of
its matter hinges on these indispensable elements. Substituting
Eq.\eqref{4-a} into Eqs.\eqref{a-1}-\eqref{a-3}, we obtain
\begin{align}\nonumber
\rho&=\bigg[e^{-\gamma_{2} r^2} \big(2 B  r^2 e^{2 \gamma_{2} r^2}
\big(A \big(r^2 (\gamma_{0}+\gamma_{2})+3\big)-3 q_{0}^2
r^4\big)+e^{\gamma_{2} r^2} \big(A \big(B \big(-\big(2 r^4
(\gamma_{0}
\\\nonumber
&-5 \gamma_{2})+r^2 (\gamma_{0}+\gamma_{2}+6)+7\big)\big)-6\big)+(7
B +6) q_{0}^2 r^4\big)+A \big(r^2 \big(-B  \big(2 \gamma_{0} r^2
\\\label{5a}
&\times(a-\gamma_{2})+\gamma_{0}+9 \gamma_{2}\big)- \gamma_{2}\big)+
B +6\big)\big)\bigg]\bigg[r^2 \big(B \big(2 (4 B +3) r^2
e^{\gamma_{2} r^2}-8 B\big)\big)\bigg]^{-1},
\\\nonumber
p_{r}&=\big[e^{-\gamma_{2} r^2} \big(e^{\gamma_{2} r^2} \big((B (3 B
+13)+6) \big(-q_{0}^2\big) r^4-A  \big(4 \gamma_{0}^2 B (4 B +3) r^6
+\gamma_{0} r^2 \big(-4
\\\nonumber
&\times(4 B +3) r^4+(B +1) (11 B +6)+2 (B (27 B +29)+6)
r^2\big)+\gamma_{2} r^2 \big((B +1)
\\\nonumber
&\times (11 B +6)+2 (B +2) (B +3) r^2\big)+B \big(13 B +2 (B +3)
r^2+17\big)+6\big)\big)
\\\nonumber
&+A \big(-2 \gamma_{0}^2 B  (3 B +2) r^4+\gamma_{0} r^2 \big(B
\big(2 \gamma_{2}(3 B +2) r^2 +21 B+43\big)+18\big)+\gamma_{2} (B
\\\nonumber
&\times  (13 B+21)+6) r^2+B  (13 B +17)+6\big)+2 r^2 e^{2 \gamma_{2}
r^2} \big(\gamma_{0} A (B +1)(11 B +6) r^2
\\\nonumber
&+ \gamma_{2} A  (B +1) (11 B +6) r^2 +B  \big(A (B +3)+(7 B +3)
q_{0}^2 r^4\big)\big)\big)\big]\big[(B +1) r^2 \big(B \big(2 (4 B
\\\label{5b}
& +3) r^2 e^{\gamma_{2} r^2}-8 B -15\big)-6\big)\big]^{-1},
\\\nonumber
p_{t}&=\big[e^{-\gamma_{2} r^2} \big(e^{\gamma_{2} r^2} \big(A B
\big(r^2 \big(2 \gamma_{0}^2 (4 B +3) r^4 +\gamma_{0} \big(2 r^2
\big(-\gamma_{2} (4 B +3) r^2+9 B +7\big)
\\\nonumber
&+B +1\big)+\gamma_{2} \big(B -2 (B +2) r^2+1\big)-2 B \big)-B
-2\big)+(B (9 B +17)+6) q_{0}^2 r^4\big)
\\\nonumber
&+A \big(-\gamma_{0}^2 (2 B +3) (3 B +2) r^4+\gamma_{0} r^2
\big(\gamma_{2} (2 B +3) (3 B +2) r^2-(3 B +4)\big)
\\\nonumber
&+\gamma_{2} (B (B +6)+6) r^2+B (B +2)\big)-2 B  r^2 e^{2 \gamma_{2}
r^2} \big(A \big((B +1) r^2 (\gamma_{0}+\gamma_{2})-B \big)
\\\label{5c}
&+(5 B +3) q_{0}^2 r^4\big)\big)\big]\big[(B +1) r^2 \big(B \big(2
(4 B +3) r^2 e^{\gamma_{2} r^2}-8 B -15\big)-6\big)\big]^{-1}.
\end{align}
\begin{figure}\center
\epsfig{file=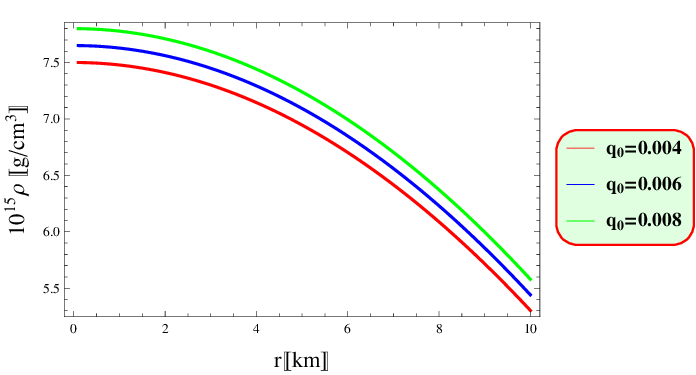,width=.4\linewidth} \epsfig{file=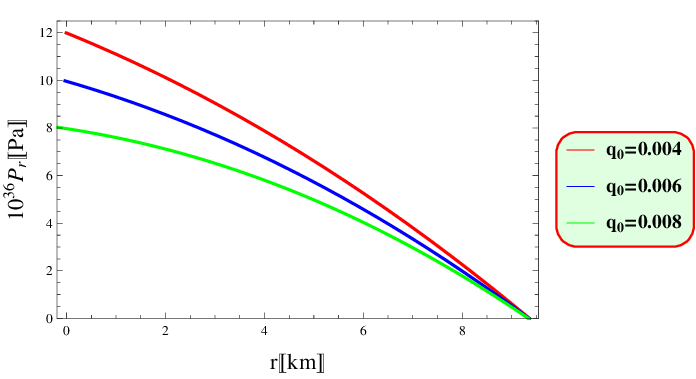,width=.4\linewidth}
\epsfig{file=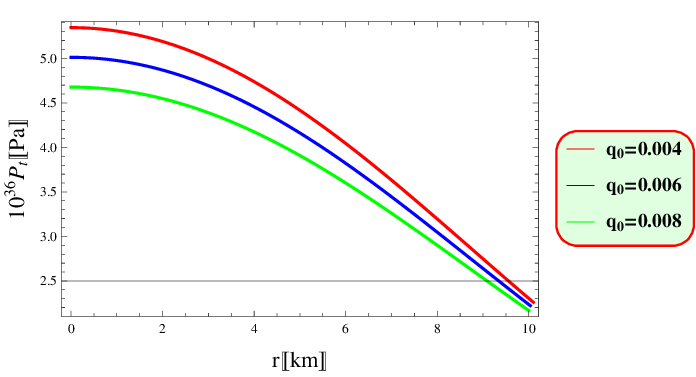, width=.4\linewidth} \caption{Plots of $\rho$, $P_{r}$, $P_{t}$
versus $r$.}
\end{figure}

We generate plots for $\rho$, $p_r$ and $p_t$ against the radial
distance to analyze the star's internal structure. For this
analysis, we set the parameters to $\gamma_{0} = 0.00363$,
$\gamma_{1} = -0.40562$ and $\gamma_{2} = 0.00405$. We choose
$\mathbf{Q} = 3$ because it provides more accurate results in the
graphs. For our graphical analysis, the values of $q_{0}$ were
varied as $0.004$, $0.006$ and $0.008$. The numerical values of the
model parameters in our study are not chosen arbitrarily but are
determined systematically based on theoretical consistency and
observational constraints. The metric parameters $\gamma_{0},
\gamma_{1}$ and $\gamma_{2}$ are obtained by matching the interior
solution to the exterior Reissner-Nordstr$\ddot{o}$m metric at the
boundary, ensuring a smooth transition between the two spacetimes.
The charge intensity parameter $q_{0}$ follows a well-established
charge function used in previous studies on charged compact objects
and is varied within a reasonable range to analyze its impact on
stability. In $f(\mathbb{Q}, \mathcal{T})$ theory, a lower charge
$q_{0}$ causes only small changes in the gravitational equations. If
$q_{0}$ stays below a certain limit, its effect on the system
remains minor, and gravity remains the dominant force. As a result,
the star's structure, stability, and density distribution are
similar to those of a neutral or weakly charged star. By comparing
cases with high and low charge, we see how different values of
$q_{0}$ affect the system. A higher charge creates stronger electric
repulsion, while a lower charge keeps gravity in control. This
balance between gravitational and electric forces plays a key role
in $f(\mathbb{Q}, \mathcal{T})$ theory, where charge significantly
influences the system behavior, especially at different distances
$r$. The coupling constant $A$ in $f(\mathbb{Q}, \mathcal{T})$
gravity is chosen based on the widely studied linear model and is
constrained to maintain a realistic mass-radius relation.
Additionally, the PS mass and radius are taken from spectroscopic
observations of PS, ensuring the astrophysical relevance of our
findings. Therefore, all parameter choices are well-motivated and
justified within the framework of the study. Figure \textbf{1}
illustrates the plots of $\rho$, $p_r$ and $p_t$. These values peak
at the star's center and gradually diminish as the radius increases,
indicating that the star has a dense core that becomes less
concentrated towards its surface, demonstrating a highly compact
profile for the PS. As the radial coordinate $r$ increases, both the
energy density and pressures gradually decrease, which is a typical
characteristic of compact stellar objects. This decreasing trend
suggests that the PS mass is primarily concentrated at the core,
while the outer layers are relatively less dense. Such a
distribution aligns with the expected structure of neutron stars,
where the core remains extremely dense and the pressure
progressively declines towards the surface.

Anisotropic fluids ($\chi$) are those in which the pressure varies
depending on the direction within the fluid. Unlike isotropic
fluids, where the pressure is uniform in all directions. Anisotropy,
expressed as $(\chi = p_{t} - p_{r})$, leads to pressure variations
that influence the system's dynamics. This characteristic is
significant in astrophysics, particularly in the modeling of compact
stars, as it influences their stability, structure and evolution
\cite{10-f}. Incorporating $\chi$ in stellar models allows for a
more accurate representation of the internal dynamics and the
effects of strong gravitational fields. The mathematical expression
for $\chi$ can be written as
\begin{align}\nonumber
\chi &=\big[e^{-\gamma_{2} r^2} \big(e^{\gamma_{2} r^2} \big((B (3 B +13)+6)
\big(-q_{0}^2\big) r^4-A  \big(4 \gamma_{0}^2 B (4 B +3) r^6+\gamma_{0} r^2 \big(-4
\\\nonumber
&\times\gamma_{2} B (4 B +3) r^4+(B +1) (11 B +6)+2 (B (27 B +29)+6)
r^2\big)+\gamma_{2} r^2 \big((B +1)
\\\nonumber
&+2 (B +2) (B +3) r^2\big)+\eta \big(13 B +2 (B +3)
r^2+17\big)+6\big)\big)+A  \big(-2 \gamma_{0}^2 B  (3 B +2)
\\\nonumber
&\times r^4+\gamma_{0} r^2 \big(B  \big(2 \gamma_{2} (3 B +2) r^2+21
B +43\big)+18\big)+\gamma_{2} (B (13 B +21)+6) r^2+B
\\\nonumber
&\times (13 B +17)+6\big)+2 r^2 e^{2 \gamma_{2} r^2} \big(\gamma_{0}
A  (B +1) (11 B +6) r^2+\gamma_{2} A (B +1) (11 B +6) r^2
\\\nonumber
&+B \big(A (B +3)+(7 B +3) q_{0}^2 r^4\big)\big)\big)\big]\big[(B
+1) r^2 \big(B \big(2 (4 B +3) r^2 e^{\gamma_{2} r^2}-8 B
-15\big)\big)\big]^{-1}
\\\nonumber
&-\big[e^{-\gamma_{2} r^2} \big(e^{\gamma_{2} r^2} \big(A B
\big(r\gamma_{2}^2 \big(2 \gamma_{0}^2 (4 B +3) r^4+\gamma_{0}
\big(2 r^2 \big(-\gamma_{2} (4 B +3) r^2+9 B +7\big)
\\\nonumber
&+B +1\big)+\gamma_{2} \big(B -2 (B +2) r^2+1\big)-2 B \big)-B
-2\big)+(B (9 B +17)+6) q_{0}^2 r^4\big)
\\\nonumber
&+A \big(-\gamma_{0}^2 (2 B +3) (3 B +2) r^4+\gamma_{0} r^2
\big(\gamma_{2} (2 B +3) (3 B +2) r^2-(3 B +4) (5 B +3)\big)
\\\nonumber
&+\gamma_{2} (B (B +6)+6) r^2+B (B +2)\big)-2 B r^2 e^{2 \gamma_{2}
r^2} \big(A \big((B +1) r^2 (a+\gamma_{2})-B \big)
\\\nonumber &+(5 B +3) q_{0}^2 r^4\big)\big)\big]\big[(B +1)
r^2 \big(B \big(2 (4 B +3) r^2 e^{\gamma_{2} r^2}-8 B
-15\big)-6\big)\big]^{-1}.
\end{align}

Figure \textbf{2} shows that the anisotropy satisfies the stability
condition. It begins at zero in the core and gradually rises as it
approaches the star's surface. This positive anisotropy plays a
crucial role in maintaining the star's stability \cite{10-gg}. This
behavior is expected because a spherically symmetric compact object
typically has isotropic pressure at the core, meaning the radial and
tangential pressures are initially equal. As we move outward, the
effects of charge, gravitational interactions and modified gravity
terms lead to a difference between these pressures, causing
anisotropy to increase. A positive anisotropy means that the
tangential pressure dominates over the radial pressure, which helps
in counteracting the inward gravitational pull and contributes to
the stability of the star. The smooth and increasing nature of the
anisotropy profile indicates that the model satisfies the necessary
stability conditions. This trend suggests that the presence of
charge influences the matter distribution in such a way that the
stellar structure remains physically viable and stable under
modified gravity.

For $q_{0} = 0.004$, the density at the core is about $\rho_{core}
\approx 9.167 \times 10^{26}$ g/cm$^{3}$. The corresponding
$p_{r(core)} \approx 10.7 \times 10^{36}$ dyn/cm$^{2}$, while
$p_{t(core)} \approx 9.36 \times 10^{36}$ dyn/cm$^{2}$. Moving to
the star's surface, the density significantly drops to $\rho_{I} =
6.851 \times 10^{26}$ g/cm$^{3}$. Here, the $p_{r(r=R)} = 0$
dyn/cm$^{2}$ and $p_{t(r=R)} \approx 4.322 \times 10^{36}$
dyn/cm$^{2}$. When considering $q_{0} = 0.006$, there is a slight
decrease in core values, where the density is $\rho_{core} \approx
9.116 \times 10^{26}$ g/cm$^{3}$. The $p_{r(core)} \approx 10.23
\times 10^{36}$ dyn/cm$^{2}$, while the $p_{t(core)} \approx 9.002
\times 10^{36}$ dyn/cm$^{2}$. At the star boundary, the density
decreases further to $\rho_{I} = 6.773 \times 10^{26}$ g/cm$^{3}$.
Here, $p_{r(r=R)}$ remains zero, and $p_{t(r=R)}$ is close to $4.157
\times 10^{36}$ dyn/cm$^{2}$. For $q_{0} = 0.008$, the core density
is around $\rho_{core} \approx 9.051 \times 10^{26}$ g/cm$^{3}$. The
pressures in this scenario are $p_{r(core)} \approx 9.479 \times
10^{36}$ dyn/cm$^{2}$ and $p_{t(core)} \approx 8.627 \times 10^{36}$
dyn/cm$^{2}$. At the surface, the density falls to $\rho_{I} = 6.709
\times 10^{26}$ g/cm$^{3}$. The pressures at this point are
$p_{r(r=R)} = 0$ dyn/cm$^{2}$ and $p_{t(r=R)} \approx 4.019 \times
10^{36}$ dyn/cm$^{2}$.
\begin{figure}\center
\epsfig{file=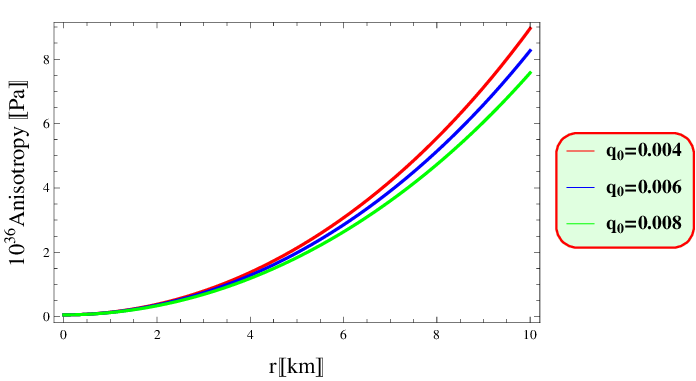,width=.5\linewidth}\caption{$\chi$ against $r$.}
\end{figure}

\subsection{Constraint on Mass-Radius Ratio}

The mass-radius ratio is a critical parameter in astrophysics that
defines the compactness of celestial objects, such as PS. It
reflects how densely matter is packed within a given radius and
plays a significant role in understanding the gravitational field
strength of these bodies. A higher mass-radius ratio indicates a
stronger gravitational pull, influencing the object's stability. The
aggregate mass of the PS can be delineated in the form of
\cite{10-g}
\begin{equation}\nonumber
M(r)= 4 \pi  \int_0^r \varsigma^2 \rho (\varsigma) d\varsigma.
\end{equation}
Figure \textbf{3} illustrates the computational solution of this
formulation, corroborating the robustness of the proposed paradigm.
The plot displays a steady and consistent growth in mass relative to
the star's radius, meeting the essential criterion that mass remains
a positive, progressively augmenting quantity. The smooth behavior
of the mass function shows that there are no sudden changes or
irregularities inside the star, making the model reliable. Most of
the mass is concentrated in the inner regions, with the function
leveling off near the surface. This pattern is typical of neutron
stars, where strong gravity compresses most of the mass into a small
space. The presence of charge $(q_0)$ affects how mass is
distributed, as seen in the different curves. A higher charge
slightly increases the mass for a given radius because the electric
force pushes outward, opposing gravity and allowing the star to
expand a little. Despite this, the model stays within astrophysical
limits, confirming that the pulsar remains stable under modified
gravity and charge effects.
\begin{figure}\center
\epsfig{file=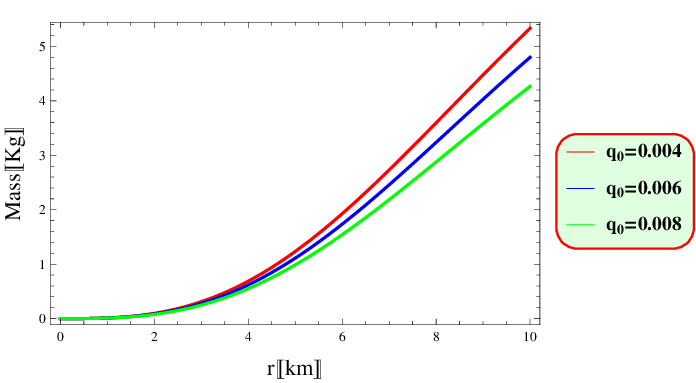,width=.48\linewidth}\caption{Mass-radius against $r$.}
\end{figure}

\subsection{Compactness}

Compactness in astrophysics refers to the ratio of an object
$\mu=\frac{M(r)}{r}$, often used to describe dense celestial bodies
like PS. It is a key parameter in determining the strength of the
gravitational field around the object, influencing phenomena such as
gravitational redshift $Z(r)$ and escape velocity. Higher
compactness indicates a stronger gravitational pull, playing a
crucial role in understanding the behavior and stability of compact
astrophysical objects. As per Buchdahl's criterion \cite{10-h}, a PS
is deemed to be achievable if its ratio, $\mu$, remains below
$\frac{4}{9}$. Figure \textbf{4} illustrates that the compactness
progressively intensifies yet consistently adheres to the stipulated
threshold of $\mu < \frac{4}{9}$, thereby ensuring the stability and
plausibility of the model. The most important result is that $\mu$
remains below the Buchdahl limit, which ensures that the PS is not
close to forming a black hole. Different curves represent varying
values of charge $( q_0 )$, showing that while charge affects the
compactness, the PS remains within stable limits for all cases. The
results confirm that the PS is stable and does not undergo
gravitational collapse. The smooth increase in compactness suggests
that the mass distribution is well-behaved and the fact that $\mu$
remains within the required limit further strengthens the validity
of the model. The presence of charge slightly alters the compactness
but does not lead to instability, reinforcing the conclusion that
the PS is a physically viable stellar structure within
$f(\mathbb{Q},\mathcal{T})$ gravity.
\begin{figure}\center
\epsfig{file=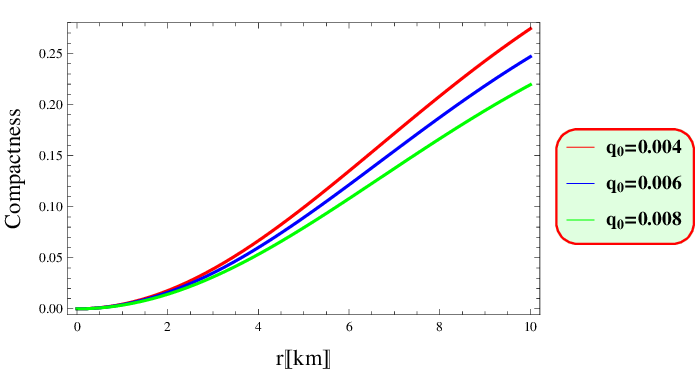,width=.48\linewidth} \caption{Compactness versus $r$.}
\end{figure}

\subsection{The Geometric Configuration}

Gravitational redshift is a phenomenon predicted by GR, where light
or electromagnetic radiation emitted from a massive object, such as
PS, is shifted to longer wavelengths as it escapes the object's
gravitational field. This occurs because the strong gravity near the
massive body affects the energy of the photons, causing them to lose
energy and stretch their wavelength as they travel away. The greater
the gravitational field, the more pronounced the redshift, serving
as an important tool for examining the characteristics and dynamics
of compact astrophysical entities. The mathematical expression for
$Z(r)$ can be formulated in the form of
\begin{equation*}
Z(r)=-1+\frac{1}{\sqrt{-g_{tt}}}.
\end{equation*}
Applying \eqref{cc-1}, the expression is reformulated into an
alternative representation
\begin{equation*}
Z(r)=-1+\frac{1}{\sqrt{1-\mu}}.
\end{equation*}
As emphasized in Ivanov's study \cite{10-i}, the anisotropic
arrangement is characterized by a value of 5.211. A graph of the
$Z(r)$ for the PS is generated using various values of $q_{0}$. As
illustrated in Figure \textbf{5} \cite{10-j}, $Z(r)$ retains a
positive and finite nature throughout the stellar body's interior,
exhibiting a gradual escalation. Notably, the redshift function
stays within the specified limit ($Z(r) < 5.211$). The gravitational
redshift describes how light or electromagnetic radiation emitted
from the pulsar surface gets shifted to longer wavelengths due to
the strong gravitational field. The plot shows that $Z(r)$ starts
from a lower value at the center and gradually increases as we move
outward. This trend is expected because the gravitational potential
is strongest near the center and weakens towards the surface,
affecting the frequency of emitted radiation accordingly. The
presence of charge slightly influences the redshift profile, with
higher charge values leading to minor variations but the overall
trend remains stable. Thus, the behavior of gravitational redshift
in this model supports the physical stability and observational
feasibility of the PS, reinforcing the compatibility of the charged
anisotropic structure with modified gravity.
\begin{figure}\center
\epsfig{file=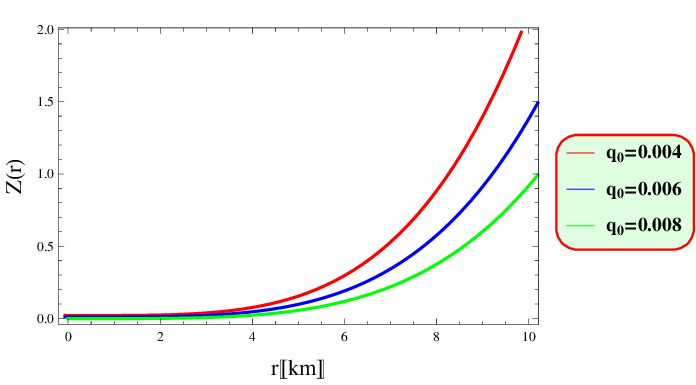,width=.48\linewidth}\caption{Redshift versus $r$.}
\end{figure}

\subsection{\textit{Energy Conditions}}

\textit{Energy conditions} are a set of inequalities in GR that
impose physical constraints on matter and energy distributions
within spacetime. In astrophysics, they play a crucial role in
evaluating the viability and stability of models, particularly in
the study of compact objects like PS.
\begin{enumerate}
\item Null \emph{energy condition} (NEC) requires that
$0 \leq \rho + p_{r},~ 0 \leq p_t+\rho$.
\item Dominant \emph{energy condition} (DEC) is expressed as
$0 \leq \rho - p_t,~ 0 \leq \rho - p_r$.
\item Weak \emph{energy condition} (WEC) can be defined as
$0 \leq \rho + p_{t},~ 0 \leq \rho + p_r, ~ 0 \leq \rho$.
\item Strong \emph{energy condition} (SEC) asserts that
$0 \leq \rho + p_t,~ 0 \leq \rho + p_r,~ 0 \leq \rho + p_r + 2p_t$.
\end{enumerate}
Energy constraints are fundamental in assessing the equilibrium and
feasibility of celestial bodies within spacetime. For an
astronomical structure to be deemed physically plausible, it must
satisfy these criteria. Figure \textbf{6} illustrates the
\emph{energy conditions} across various $q_{0}$ values. \emph{Energy
conditions} are crucial in GR and modified gravity theories to
ensure the physical viability of a stellar model. The plots verify
whether the PS satisfies the necessary conditions for a stable and
physically reasonable matter distribution. The figure consists of
five graphs corresponding to different \emph{energy conditions}. The
NEC show that these conditions hold throughout the stellar interior,
meaning that the total energy density combined with pressure remains
non-negative. The DEC ensure that the energy density is always
greater than or equal to the radial and tangential pressures. The
figure confirms that DEC is satisfied, meaning that matter within
the pulsar behaves like a realistic fluid. The WEC is crucial for
ensuring that the energy density remains positive throughout the
star. The figure shows that WEC holds for all values of $r$, further
supporting the model physical consistency. The SEC is important for
determining whether the gravitational field behaves normally within
the modified gravity framework. The figure demonstrates that SEC is
satisfied, meaning that the star self-gravity is strong enough to
hold the structure together. The smooth nature of the curves
indicates that there are no singularities or unphysical regions
within the stellar interior. This reinforces the stability and
viability of the charged PS model in $f(\mathbb{Q},\mathcal{T})$
gravity.
\begin{figure}\center
\epsfig{file=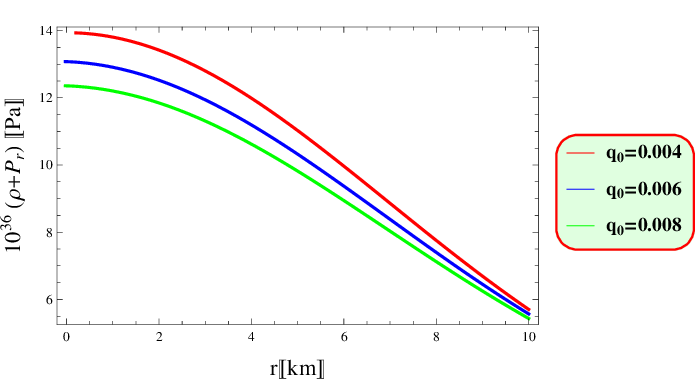,width=.4\linewidth} \epsfig{file=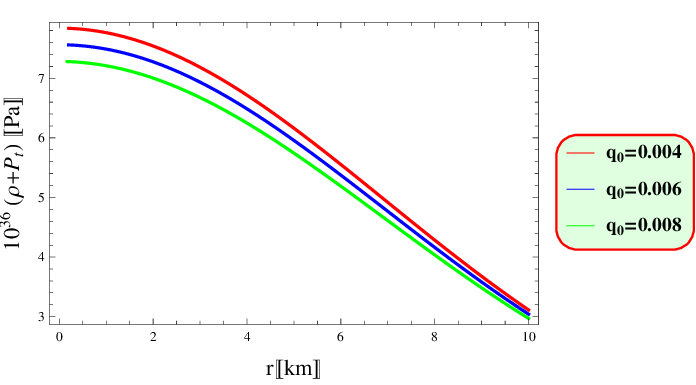,width=.4\linewidth}
\epsfig{file=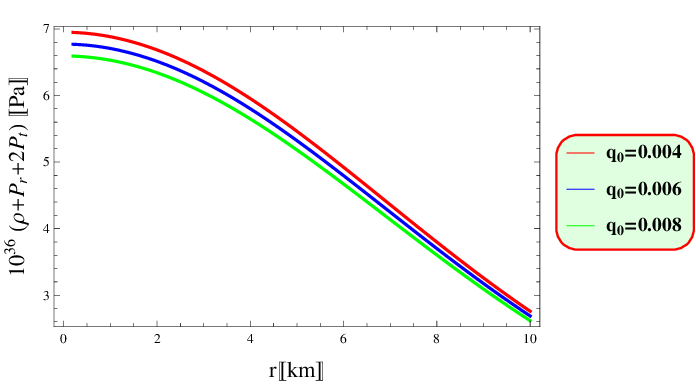,width=.4\linewidth} \epsfig{file=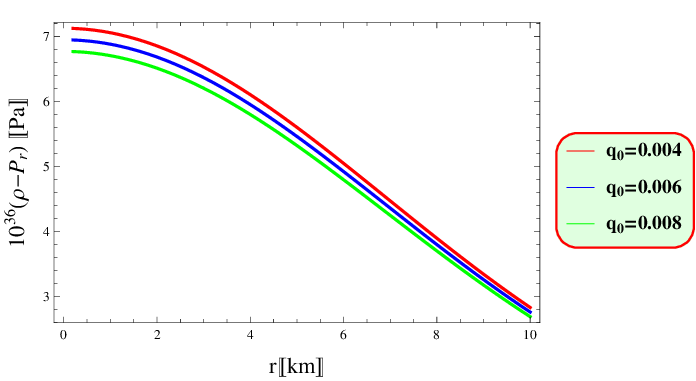,width=.4\linewidth}
\epsfig{file=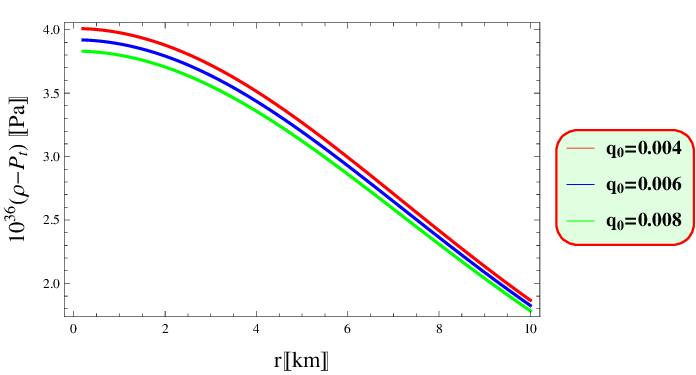,width=.4\linewidth} \caption{Plots of \emph{energy conditions}
versus $r$.}
\end{figure}

\subsection{The Zeldovich Condition}

The Zeldovich Condition \cite{10-k} is a key criterion in
astrophysics that imposes an upper limit on the equation of state
(EoS) parameter to ensure physical viability in dense stellar
objects. It posits that the ratio $\frac{p(0)}{\rho(0)}$ must remain
less than or equal to 1, where $\rho(0)$ signifies the core's
density and $p(0)$ denotes the pressure at the stellar core. This
condition ensures that the speed of sound within the star remains
below the speed of light, thereby upholding causality and stability
in compact objects like neutron stars. To verify this, we undertake
an evaluation of $\rho$, $p_r$ and $p_t$ as $r$ approaches to zero.
The computed values are presented below
\begin{align}\nonumber
\rho(0)&=\frac{A \big(-2\gamma_{0}B - 17B \gamma_{2} -
18\gamma_{2}\big)}{-8B^2 - 15B - 6},
\\\nonumber
p_{r}(0)&=\frac{A \big(10\gamma_{0}B^2 + 26\gamma_{0}B +
12\gamma_{0} - 11B^2\gamma_{2} - 13B \gamma_{2} -
6\gamma_{2}\big)}{(B + 1)(-8B^2 - 15B - 6)},
\\\nonumber
p_{t}(0)&=\frac{A \big(-14\gamma_{0}B^2 - 28\gamma_{0}B -
12\gamma_{0} + 2B^2\gamma_{2} + 7B \gamma_{2} + 6\gamma_{2}\big)}{(B
+ 1)(-8B^2 - 15B - 6)}.
\end{align}
Utilizing the previously determined values for the PS, we are able
to scrutinize this further. These calculations reveal that
$\frac{p_r(0)}{\rho(0)} = -0.0842$, which is clearly less than 1.
Similarly, $\frac{p_t(0)}{\rho(0)} = 0.3404$, which also meets the
condition of being less than 1. These results validate the Zeldovich
condition, confirming the model's stability and theoretical
soundness. This validation is essential, as it establishes that the
core $\frac{p(0)}{\rho(0)}$ complies with the required physical
limitations, thereby bolstering the framework's comprehensive
legitimacy.

\subsection{Causality Conditions}

This criterion dictates that no signal can surpass the velocity of
light, ensuring that the temporal interval between any two events
within spacetime remains non-negative. This criterion is crucial for
maintaining the model's validity. Thus, the causality constraint is
crucial for guaranteeing the stability and consistency of models
dealing with anisotropic fluids in the realm of relativistic
astrophysics. For such fluids, this constraint specifies that
\begin{equation}\label{11}
v_{r}^{2}= \frac{dp_r}{d\rho},\quad v_{t}^{2}= \frac{dp_t}{d\rho}.
\end{equation}
For PS, maintaining structural stability necessitates that the
squared velocity of sound is constrained to the range $(0, 1)$
\cite{10-l}. Both the $v_{r}^{2}$ and $v_{t}^{2}$ meet this
criterion, where $0 \leq v_r^2 \leq 1$ and $0 \leq v_t^2 \leq 1$.
The pulsar's stability can also be evaluated using the cracking
condition, which requires $-1 < |v_t^2 - v_r^2| < 0$ \cite{10-m}.
The derivation of these acoustic velocity components is detailed in
Appendix \textbf{A}. Figure \textbf{7} depicts the variation of
$v_{r}^{2}$ and $v_{t}^{2}$ across the radial coordinate. The
recorded intervals are $0.79 < v_r^2 < 0.73$ and $0.85 < v_t^2 <
0.72$. Throughout the entirety of the charged PS's interior, the
inequality $0.0120 < |v_t^2 - v_r^2| < 0.0116$ is perpetually
upheld, thereby affirming the robustness of its anisotropic
characteristics.

These conditions ensure that the model remains physically consistent
by verifying whether the speed of sound remains within acceptable
limits and whether the PS satisfies the cracking condition for
stability. The figure consists of three graphs, each showing
different aspects of the stability conditions. The first graph
displays $v_{r}^{2}$, which represents the speed at which radial
pressure perturbations propagate within the pulsar. The plot shows
that $v_{r}^{2}$ remains between 0 and 1 throughout the stellar
interior, ensuring that the model satisfies the causality condition,
meaning that no signals travel faster than the speed of light. The
second graph illustrates $v_{t}^{2}$, which represents the speed of
tangential pressure perturbations. Similar to $v_{r}^{2}$, this
quantity also lies within the range $ 0 \leq v_{t}^{2} \leq 1 $,
confirming that the pulsar structure does not allow any superluminal
(faster-than-light) signal propagation. The third graph shows the
difference $ v_{t}^{2} - v_{r}^{2}$, which is crucial for
determining stability against local perturbations. The condition $0
\leq \mid v_{t}^{2} - v_{r}^{2} \mid \leq 1$ must hold for a stable
stellar configuration. The plot confirms that this difference
remains positive and within the required range, implying that no
sudden instability (cracking) occurs within the PS. The results in
Figure \textbf{7} strongly indicate that the PS satisfies the
necessary stability and causality conditions in
$f(\mathbb{Q},\mathcal{T})$ gravity. The smooth and well-behaved
nature of the plots suggests that the pulsar remains dynamically
stable under small perturbations. Moreover, the influence of charge
$q_{0}$ slightly affects the numerical values but does not disturb
the overall stability.
\begin{figure}\center
\epsfig{file=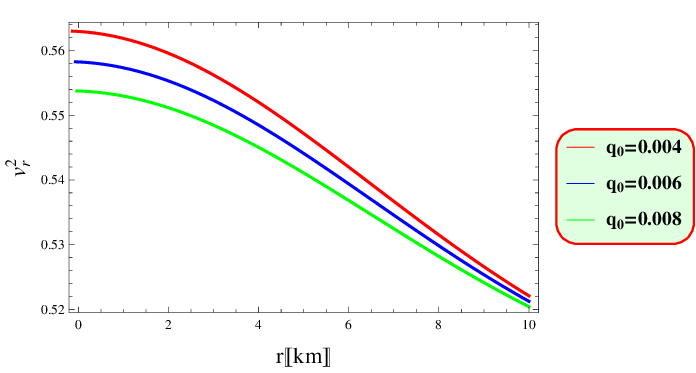,width=.4\linewidth} \epsfig{file=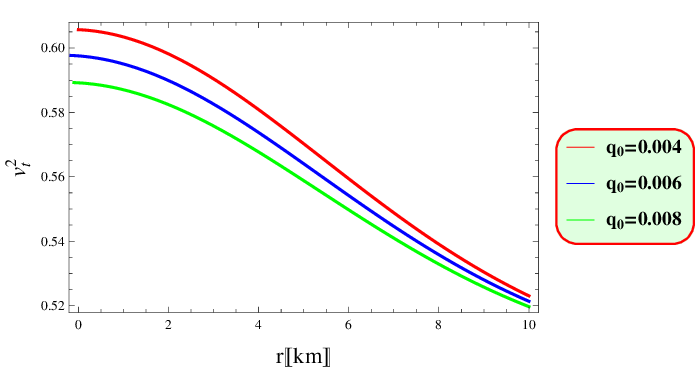,width=.4\linewidth}
\epsfig{file=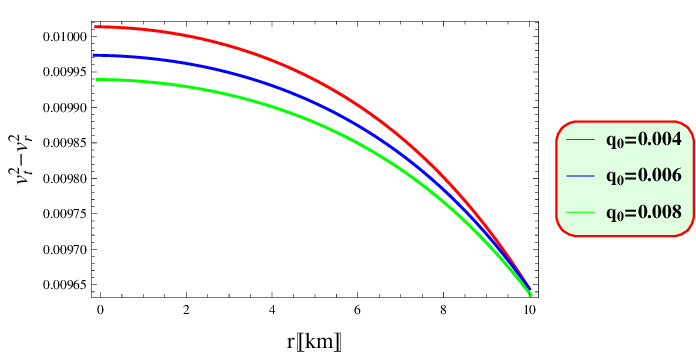,width=.4\linewidth} \caption{Plots of causality conditions
versus $r$.}
\end{figure}

\section{The EoS Parameter}

Under this framework, we apply the ensuing interrelationships
\cite{7-e}
\begin{align}\label{E}
p_{r}(\rho) \approx v^2_{r} (\rho - \rho_I), \quad p_{t}(\rho)
\approx v^{2}_{t} (\rho - \rho_{II}),
\end{align}
In this scenario, $\rho_I$ signifies the $\rho$ at which the $p_r$,
vanishes. Both $\rho_I$ and $\rho_{II}$ represent the surface
densities of the star, associated with the $p_r$ and $p_t$,
correspondingly. However, it is important to emphasize that while
$\rho_I$ leads to zero $p_r$, $\rho_{II}$ does not inherently result
in $p_t = 0$. By employing this model, precise quantifications of
both the acoustic propagation velocity and the surface $\rho$ are
meticulously attained. As a case in point, when $q_{0} = 0.004$,
using Eqs.\eqref{38} and \eqref{39}, the values obtained are $v_r^2
\approx 0.71$, $v_t^2 \approx 0.82$, $\rho_I = 5 \times 10^{26}$
g/cm$^3$, and $\rho_{II} \approx 4.2 \times 10^{26}$ g/cm$^3$. For
$q_{0} = 0.006$, the calculations yield $v_r^2 \approx 0.72$, $v_t^2
\approx 0.83$, $\rho_I = 5.2 \times 10^{26}$ g/cm$^3$, and
$\rho_{II} \approx 4.3 \times 10^{26}$ g/cm$^3$. Similarly, when
$q_{0} = 0.008$, the results are $v_r^2 \approx 0.787$, $v_t^2
\approx 0.839$, $\rho_I = 5.9 \times 10^{26}$ g/cm$^3$, and
$\rho_{II} \approx 4.0 \times 10^{26}$ g/cm$^3$. Within myriad
physical systems, the EoS parameter is indispensable for elucidating
the intricate interrelation between $p$ and $\rho$. To deem a model
physically tenable, it is imperative that the $\omega_{r}$ and
$\omega_{t}$ are confined within the interval $(0,1)$ \cite{10-n}.
This correlation is delineated as
\begin{equation}\label{C}
\omega_{r}= \frac{p_{r}}{\rho},\quad \omega_{t}= \frac{p_{t}}{\rho}.
\end{equation}
Inserting Eqs.\eqref{5a}-\eqref{5c} into \eqref{C}, we obtain
\begin{align}\nonumber
\omega_{r}&=\big[e^{\gamma_{2} r^2} \big((B  (3 B +13)+6)
\big(-q_{0}^2\big) r^4-A  \big(4 a^2 B  (4 B +3) r^6+\gamma_{0} r^2
\big(-4 \gamma_{2} B
\\\nonumber
&\times(4 B +3) r^4+(1+B) (B +6)+2 (B  (27 B)+6) r^2\big)+\gamma_{2}
r^2 \big((B +1)
\\\nonumber
&\times(B +6)+2 (2+B) (3+B) r^2\big)+\big(13 B +(B +3)
r^2+17\big)+6\big)\big)
\\\nonumber
&+\big(-2 a^2 B  (3 B +2) r^4+\gamma_{0} r^2 \big(B \big(2
\gamma_{2} (3 B +2) r^2+21 B +43\big)+18\big)
\\\nonumber
&+\gamma_{2} (B  (13 B +21)+6) r^2+B (13 B +17)+6\big)+2 r^2 e^{2
\gamma_{2} r^2} \big(\gamma_{0} A (B +1)
\\\nonumber
&\times (11 B +6) r^2)+\gamma_{2} A (B +1) (11 B +6) r^2+B \big(A (B
+3+(7 B +3) q_{0}^2 r^4\big)\big)\big]
\\\nonumber
&\times\big[(B +1) \big(2 B r^2 e^{2 \gamma_{2} r^2} \big(A \big(r^2
(\gamma_{0} +\gamma_{2})+3\big)-3 q_{0}^2 r^4\big)+e^{\gamma_{2}
r^2} \big((7 B +6) q_{0}^2 r^4
\\\nonumber
&-A \big(B \big(2 r^4 (\gamma_{0} -5 \gamma_{2})+r^2 (\gamma_{0}
+\gamma_{2}+6)+7\big)+6\big)\big)A \big(r^2 \big(-B \big(2
\gamma_{0}r^2 (\gamma_{0} -\gamma_{2})
\\\nonumber
&+\gamma_{0} +9 \gamma_{2}\big)-12 \gamma_{2}\big)+7 B
+6\big)\big)\big]^{-1},
\\\nonumber
\omega_{t}&=\big[e^{\gamma_{2} r^2} \big(A  B  \big(r^2 \big(2 a^2
(4 B +3) r^4+\gamma_{0} \big(-2 \gamma_{2} (4 B +3) r^4+B +2 (9 B
+7) r^2+1\big)
\\\nonumber
&+\gamma_{2} \big(-2+B (2+B) r^2+1\big)-2 B \big)-2-B\big)+(B (9 B
+17)+6) q_{0}^2 r^4\big)+A \big(-a^2
\\\nonumber
&\times(2 B +3) (3 B +2) r^4+\gamma_{0} r^2 \big(\gamma_{2} (2 B +3)
(3 B +2) r^2-(3 B +4) (5 B +3)\big)
\\\nonumber
&+\gamma_{2} (B (B +6)+6) r^2+B (B +2)\big)-2 B  r^2 e^{2 \gamma_{2}
r^2} \big(A \big((B +1) r^2 (a+\gamma_{2})-B \big)
\\\nonumber
&+(5 B +3) q_{0}^2 r^4\big)\big]\big[(B +1) \big(2 B  r^2 e^{2
\gamma_{2} r^2} \big(A \big(r^2 (\gamma_{0}+\gamma_{2})+3\big)-3
q_{0}^2 r^4\big)
\\\nonumber
&+e^{\gamma_{2} r^2} \big((7 B +6) q_{0}^2 r^4-A \big(B \big(2 r^4
(\gamma_{0}-5 \gamma_{2})+r^2 (\gamma_{0}+\gamma_{2}+6)
\\\nonumber
&+7\big)+6\big)\big)+A \big(r^2 \big(-B \big(2 \gamma_{0} r^2
(\gamma_{0}-\gamma_{2})+\gamma_{0}+9 \gamma_{2}\big)-12
\gamma_{2}\big)+7 B +6\big)\big)\big].
\end{align}

Figure \textbf{8} presents the optimal EoS for the PS, depicting the
correlation between $\rho$ and $p_{r}$ across distinct quantities of
$q_{0}$. The findings adhere to a EoS pattern, satisfying the
prerequisites for a plausible PS configuration. Likewise, the
tangential EoS exhibits a strong alignment with a linear model.
These deduced EoSs, exhibiting particular validity near the stellar
centroid, permeate the entirety of its interior domain confined
within the interval $(0,1)$. The first two graphs show the variation
of $p_{r}$ and $p_{t}$ as functions of $\rho$. These plots confirm
that the pressures follow a linear relationship with density,
supporting the assumption of a realistic EoS. The third graph
illustrates the behavior of the $\omega_{r}$ remains within the
range $ 0 \leq \omega_r \leq 1 $ throughout the stellar interior,
which is a necessary condition for a physically viable stellar
model. The fourth graph shows the $\omega_{t}$ also stays positive
and finite, ensuring the stability of the model. These results
confirm that the PS follows a physically reasonable EoS, where the
pressure remains proportional to the energy density.
\begin{figure}\center
\epsfig{file=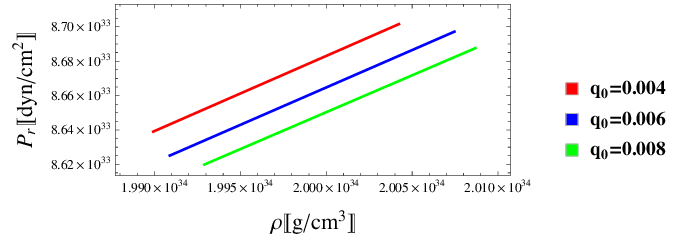,width=.48\linewidth}
\epsfig{file=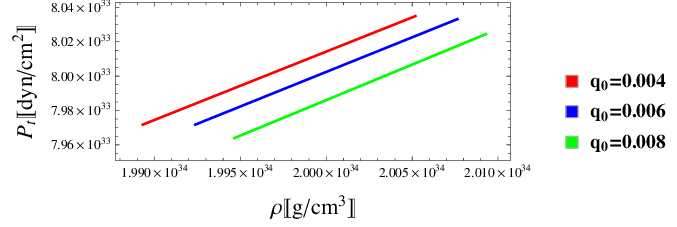,width=.48\linewidth}
\epsfig{file=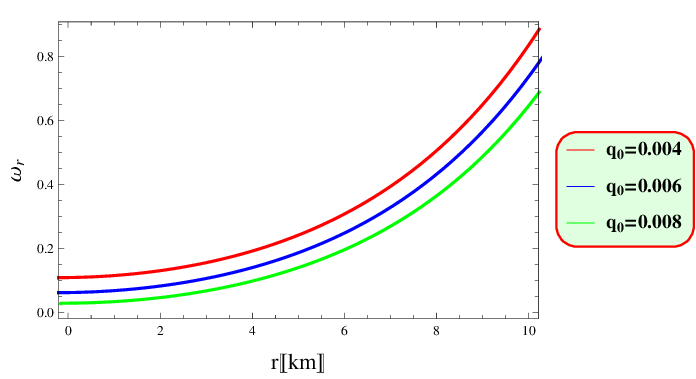,width=.48\linewidth}
\epsfig{file=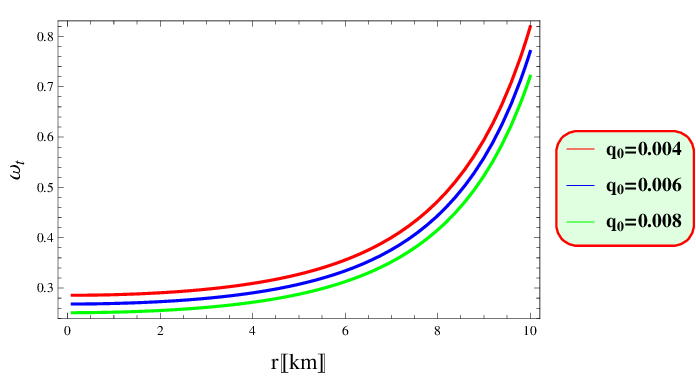,width=.48\linewidth} \caption{EoS versus $r$.}
\end{figure}

\subsection{Adiabatic Index and Hydrodynamic Stability}

An indispensable parameter for ascertaining the equilibrium of the
PS is the adiabatic index $\Gamma$, which is delineated as
\cite{10-p}
\begin{equation*}
\Gamma=\frac{4}{3}\bigg(1+\frac{\sigma}{r|
\acute{p}_{r}|}\bigg)_{max},
\end{equation*}
\begin{equation}\label{aa}
\Gamma_{r}=\frac{p_{r}+\rho}{p_{r}}v^{2}_{r},~~\Gamma_{t}
=\frac{p_{t}+\rho}{p_{t}}v^{2}_{t}.
\end{equation}
In this context, $\Gamma_{t}$ and $\Gamma_{r}$ represent the
tangential and radial components of the $\Gamma$. Under isotropic
conditions $(\chi = 0)$, the $\Gamma$ converges to its canonical
value of $\frac{4}{3}$. Conversely, the introduction of mild $(\chi
\neq 0)$ induces a diminution of the index below $\frac{4}{3}$,
thereby aligning with the established tenets of dynamical
equilibrium \cite{10-q}. Appendix \textbf{B} elaborately delineates
the detailed mathematical expressions pertaining to $\Gamma$. Figure
\textbf{9} shows that in $f(\mathbb{Q}, \mathcal{T})$ gravity, the
condition $\Gamma > \frac{4}{3}$ is met. This affirms the stability
of the anisotropic configuration of the PS across different values
of $q_{0}$. This indicates that increasing charge slightly affects
the stability but does not violate the fundamental stability
condition. The results confirm that the PS remains physically viable
and structurally stable, reinforcing that the inclusion of charge
within $f(\mathbb{Q}, \mathcal{T})$ gravity contributes to
maintaining equilibrium without leading to instability.
\begin{figure}\center
\epsfig{file=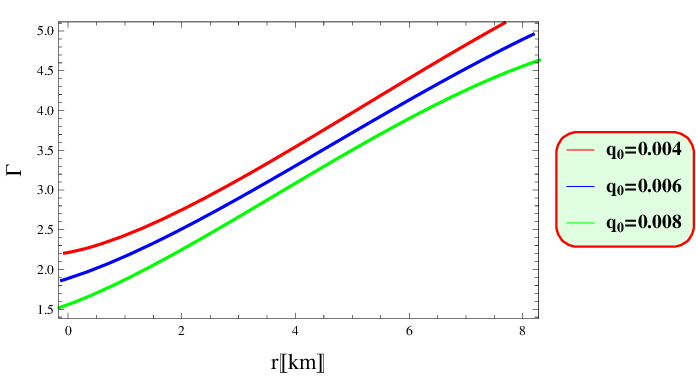,width=.48\linewidth}
\epsfig{file=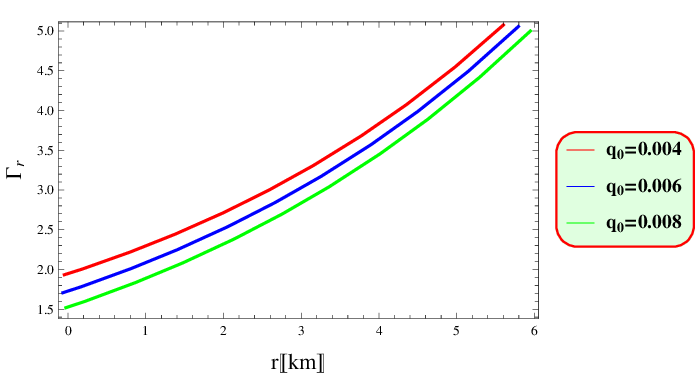,width=.48\linewidth}
\epsfig{file=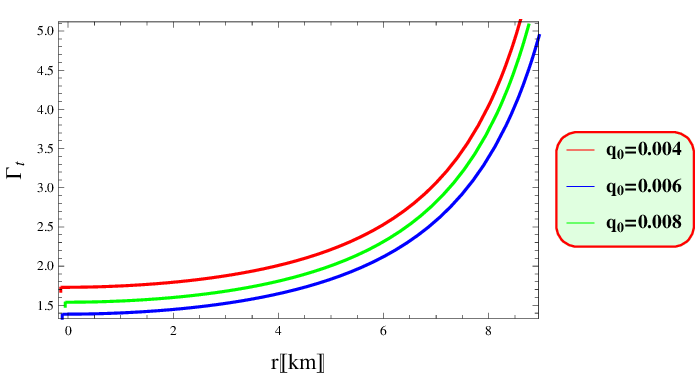,width=.48\linewidth} \caption{Adiabatic index
against $r$.}
\end{figure}

The Tolman-Oppenheimer-Volkoff (TOV) equation is a fundamental
formulation that defines the conditions necessary for equilibrium in
a static spherically symmetric stellar structure. It encapsulates
how the star's intrinsic pressure counterbalances gravitational
forces, thereby maintaining its structural stability. Such an
equation is indispensable for analyzing the structure and dynamics
of pulsars. By undertaking a meticulous analytical examination of
the TOV equation, researchers can elucidate the intrinsic $p$,
$\rho$ and appraise the overarching stability of such compact
stellar entities. Articulated for an anisotropically characterized
matter assemblage, the TOV equation is expressed in the ensuing
formulation \cite{10-r}
\begin{equation}\label{d}
\frac{dp_r}{dr}+\frac{1}{r^2}e^{\frac{\alpha-\beta}{2}}\mathbf{M}_{G}(r)
(\rho + p_r) -\frac{2}{r} (p_t - p_r) = 0,
\end{equation}
where $\mathbf{M}_{G}$ is defined as \cite{10-s}
\begin{equation}\label{d1}
\mathbf{M}_{G}(r) = 4\pi \int (\mathcal{T}^t_t- \mathcal{T}^r_r
-\mathcal{T}^\phi_\phi-\mathcal{T}^\theta_\theta) r^2
e^{\frac{\alpha+\beta}{2}} dr.
\end{equation}
Upon solving this equation, we arrive at
\begin{equation}\label{d2}
\mathbf{M}_{G}(r) = \frac{1}{2} r^2 e^{\frac{\beta-\alpha}{2}}
\alpha^{\prime}.
\end{equation}
Inserting Eq.\eqref{d2} into \eqref{d}, we get
\begin{equation}\label{33}
\frac{1}{2} \alpha^{\prime} (\rho + p_r) + \frac{dp_r}{dr} -
\frac{2}{r} (p_{t} - p_r) = 0.
\end{equation}

Employing a modified TOV formulation that assimilates $f(\mathbb{Q},
\mathcal{T})$ theory, we delve into the analysis of the hydrodynamic
equilibrium of the proposed model
\begin{equation}
F_{g}+F_{a}+F_{h}+F_{(\mathbb{Q}, \mathcal{T})}=0.
\end{equation}
In this context, the forces $F_g$, $F_a$ and $F_h$ correspond, in
respective order, to the gravitational, anisotropic and hydrostatic
forces, augmented by $F_{(\mathbb{Q}, \mathcal{T})}$. These are
defined as follows
\begin{eqnarray}\nonumber
F_{g}&=& \frac{(\rho+p_{r})\alpha^{\prime}}{2},\quad F_{a}=
\frac{2\sigma}{r},\\
\nonumber F_{h}&=& -p_{r}^{\prime},\quad
F_{(\mathbb{Q},\mathcal{T})}=
p_{r}^{\prime}+\frac{1}{2}(\rho-p_{r})\beta^{\prime}(r)-2r^{-1}(p_{t}-p_{r}).
\end{eqnarray}
After simplifying these values, we get
\begin{eqnarray}\nonumber
F_{a}&=& -\big[6 e^{-\gamma_{2} r^2} \big(e^{\gamma_{2} r^2} \big(-A
\big(2 \gamma_{0}^2 B  (4 B +3) r^6 +2 \gamma_{0} r^2
\big(-\gamma_{2} B (4 B +3) r^4+2 B ^2
\\\nonumber
&+&3 B +2 \big(6 B ^2+6 B +1\big) r^2+1\big)+2 \gamma_{2} r^2 \big(2
B ^2+3 B +(\eta +2) r^2+1\big)+4 B ^2
\\\nonumber
&+&5 B+2 B r^2+2\big)-2 \big(2 \eta ^2+5 B +2\big) q_{0}^2
r^4\big)+A \big(\gamma_{0}^2 (3 B +2) r^4+\gamma_{0} r^2 \big(2
\big(6 \eta ^2
\\\nonumber
&+&12 B +5\big)-\gamma_{2} (3 B +2) r^2\big)+4 \eta ^2
\big(\gamma_{2} r^2+1\big)+5 B \big(\gamma_{2} r^2+1\big)+2\big)+2
r^2
\\\nonumber
&\times&e^{2\gamma_{2} r^2}\big(2 \gamma_{0} A \big(2 \eta ^2+3 B
+1\big) r^2+2 \gamma_{2} A \big(2 B ^2+3 B +1\big) r^2+B \big(A +2
(2 B +1)
\\\nonumber
&\times&q_{0}^2 r^4\big)\big)\big)\big]\big[(B +1) r^3 \big(8 B ^2
\big(r^2 e^{\gamma_{2} r^2}-1\big)+3 B \big(2 r^2 e^{\gamma_{2}
r^2}-5\big)-6\big)\big]^{-1},
\end{eqnarray}
\begin{eqnarray}\nonumber
F_{g}&=&-\big[\gamma_{0} e^{-\gamma_{2} r^2} \big(A
\big(e^{\gamma_{2} r^2} \big(2 \gamma_{0}^2 B  (4 B +3)
r^6+\gamma_{0} r^2 \big(-2 \gamma_{2} B (4 B +3) r^4+6 B ^2
\\\nonumber
&+&9 B+\big(28 B ^2+30 B +6\big) r^2+3\big)+\gamma_{2} r^2 \big(6 B
^2+9 B +\big(6-4 \eta ^2\big) r^2+3\big)
\\\nonumber
&+&10 B ^2+15 B+4 B ^2 r^2+6 B r^2+6\big)+\gamma_{0}^2 B (4 B +3)
r^4-\gamma_{0} r^2 \big(2 B ^2 \big(2 \gamma_{2} r^2
\\\nonumber
&+&5\big)+3 \eta \big(\gamma_{2} r^2+7\big)+9\big)-2 r^2 e^{2
\gamma_{2} r^2} \big(3 \gamma_{0} \big(2 B ^2+3 \eta +1\big) r^2+3
\gamma_{2} \big(2 \eta ^2
\\\nonumber
&+&3 B +1\big) r^2+B (2 B +3)\big)-2 \gamma_{2} B ^2 r^2+3
\gamma_{2} r^2-10 B ^2-15 B -6\big)-2 B ^2
\\\nonumber
&+&q_{0}^2 r^4 e^{\gamma_{2} r^2} \big(r^2 e^{\gamma_{2}
r^2}+1\big)\big)\big]\big[(B +1) r\big( \eta ^2 \big(r^2
e^{\gamma_{2} r^2}-1\big)+\eta \big(2 r^2 e^{\gamma_{2}
r^2}\big)\big)\big]^{-1},
\end{eqnarray}
\begin{eqnarray}\nonumber
F_{h}&=&\bigg[2 e^{-\gamma_{0} r^2} \big(-4 B  r^2 e^{2 \gamma_{0}
r^2} \big(A  \big(-\gamma_{2}^2 \eta (4 B +3)^2 r^6 \big(\gamma_{0}
r^2-\big)+\big(\gamma_{0}^2 \eta (4 B +3)^2
\\\nonumber
&\times&\gamma_{2} r^2 r^6+\gamma_{0} r^2 \big(\eta ^3 \big(2-52
r^2\big)+B ^2 \big(8-79 r^2\big)+B \big(9-30 r^2\big)+3\big)
\\\nonumber
&+&\big(2 B ^3+8 \eta ^2 \gamma_{2}+9 B +3\big)\big)+
\big(\gamma_{0}^2 r^4 \big(B ^3 \big(4 r^2+2\big)+B ^2 \big(11
r^2+8\big)
\\\nonumber
&+&\eta\big(6 r^2+9\big)+3\big)+\gamma_{0} r^2 \big(9 B+4 B ^3 r^2+3
B ^2 \big(r^2+2\big)+3\big)+B (4 B +3)
\\\nonumber
&\times&\big(B +B r^2+2\big)\big)\big)+q_{0}^2 r^4\big(2 \gamma_{0}
B ^2 (B +1) r^2+\big(40 B ^3+99 B ^2+75 B
\\\nonumber
&+&18\big)\big)\big)+e^{\gamma_{0} r^2} \big(A B \big(-4
\gamma_{2}^2 (4 B +3) r^6 \big(\gamma_{0} \big(6 B ^2+13 B +6\big)
r^2-\big(8 B ^2
\\\nonumber
&+&15 B +6\big)\big)+4 \gamma_{2} r^4 \big(\gamma_{0}^2 \big(24 B
^3+70 B ^2+63 B +18\big) r^4-\gamma_{0} \big(92 B ^3+245 B ^2
\\\nonumber
&+&204 B +54\big)r^2+3 (B +1)^2 (2 B +1)\big)+ \big(4 \gamma_{0}^2
\big(4 B ^3+27 B ^2+42 B\big) r^6
\\\nonumber
&+&4 \gamma_{2} \big(2 B ^3+9 B ^2+9 B +3\big) r^4+(B +2)\big(8 B ^2
\big(2 r^2+1\big)+3 B \big(4 r^2+5\big)
\\\nonumber
&+&6\big)\big)\big)+\big(72 B ^4+271 B ^3+357 B ^2+192 B +36\big)
q_{0}^2 r^4\big)+A \big(8 B ^2+15 B +6\big)
\\\nonumber
&\times&\big(\gamma_{2}^2 \big(6 B ^2+13 B +6\big) r^4
\big(\gamma_{2} r^2-\big)+\gamma_{2} \gamma_{0} r^4 \big(3 \big(7 B
^2+14 B +6\big)-\gamma_{0} \big(6 B ^2
\\\nonumber
&+&13 B +6\big) r^2\big)-\big(\gamma_{0}^2 \big(B ^2+6 B +6\big)
r^4+\gamma_{0} B (B +2) r^2+B (B +2) R^4\big)\big)
\\\nonumber
&+&4 B ^2 (4 B +3) r^4e^{3 \gamma_{0} r^2}\big(A B +(5 B +3) q_{0}^2
r^4\big)\big)\bigg]\bigg[(B +1) r^3 \big(8 B ^2 \big(r^2
e^{\gamma_{0} r^2}\big)
\\\nonumber
&+&3 B \big(2 r^2 e^{\gamma_{0} r^2}-5\big)-6\big)^2\bigg]^{-1}.
\end{eqnarray}

We demonstrate the condition for stable equilibrium using a
graphical method. Figure \textbf{10} illustrates that equilibrium is
reached when the combined forces $F_{(\mathbb{Q}, \mathcal{T})}$,
$F_{a}$, $F_{g}$ and $F_{h}$ balance out to zero. This outcome
indicates that the PS model maintains stability across different
values of $q_{0}$. The figure consists of two separate graphs
showing the behavior of $F_{(\mathbb{Q}, \mathcal{T})}$. In one
graph, where $F_{(\mathbb{Q}, \mathcal{T})} = 0$, there is no
electromagnetic force, meaning the system stability relies solely on
the equilibrium between gravitational, hydrostatic, and anisotropic
forces. If these forces are well-balanced, stability can be
achieved, although the absence of an electromagnetic force might
lead to changes in pressure and structure. The other graph presents
the scenario where $F_{(\mathbb{Q}, \mathcal{T})} \neq 0$, in which
the electromagnetic force introduces additional repulsion, helping
to counter gravitational collapse and thus playing a crucial role in
sustaining the system's stability.
\begin{figure}\center
\epsfig{file=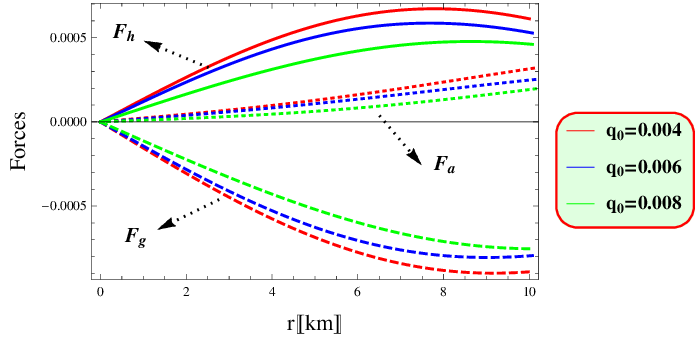,width=.48\linewidth}
\epsfig{file=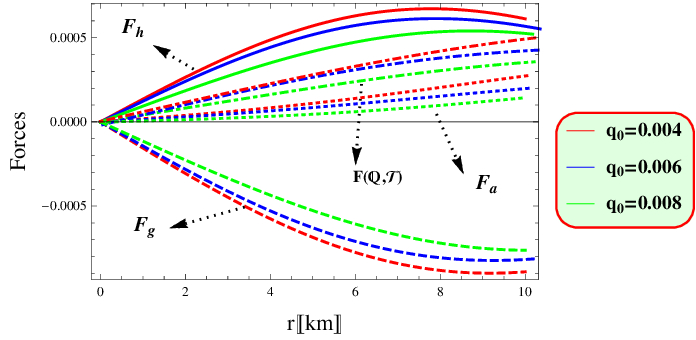,width=.48\linewidth}\caption{TOV equation as a
function of $r$.}
\end{figure}

\section{Summary}

This study delves into the effects of the $f(\mathbb{Q},
\mathcal{T}) = A\mathbb{Q} + B\mathcal{T}$ model on the structural
dynamics and stability of charged PS. By introducing an anisotropic
fluid distribution and deploying the KB ansatz to characterize the
PS internal geometry, we derive pivotal constraints on the model
parameters, shedding light on the intricate interplay between the
gravitational theory and the stellar configuration. We also explore
how the charge intensity $q_0$ affects our model through both
numerical analyses and graphical representations. The main findings
are summarized as follows
\begin{itemize}
\item
Our analysis reveals that the $\rho$, along with both $p_{r}$ and
$p_{t}$, peaks at the star's core and progressively decreases
towards the surface. Notably, the $p_{r}$ becomes zero at the star's
surface. With an increase in charge intensity, both the energy
density and pressure decrease. Despite these changes, the overall
decreasing trend of these profiles persists within this model
(Figure \textbf{1}).
\item
At the star's center, the anisotropy is zero because the $p_t$ equals the $p_r$.
However, as we move outward from the center, the $\chi$ becomes positive (Figure
\textbf{2}). This behavior aligns with theoretical predictions, thus confirming the
proposed stellar configuration's empirical viability.
\item As the radius of the PS expands, its mass grows consistently and
uniformly (Figure \textbf{3}). This trend indicates the stability of the PS.
\item
The star's compactness gradually intensifies across its interior
while remaining constrained by the limit of $\mu < \frac{4}{9}$ for
different values of $q_{0}$ (Figure \textbf{4}).
\item
By persistently maintaining values under the demarcated boundary
$Z(r) < 5.211$ whilst steadily increasing, the redshift function
comports with theoretical predictions (Figure \textbf{5}).
\item
Analysis of the \textit{energy conditions} over a spectrum of of
$q_{0}$ values demonstrates that the PS model adheres to all
essential energy constraints. (Figure \textbf{6}). This verifies the
physical viability of our model.
\item
For various values of $q_{0}$, it is found that the expressions $(0 \leq v_r^2 \leq
1)$ and $(0 \leq v_t^2 \leq 1)$ are satisfied (Figure \textbf{7}), confirming the
stability and causality requirements. Additionally, our analysis shows that the
condition $(0 \leq |v_t^2 - v_r^2| \leq 1)$ is satisfied throughout the pulsar's
interior, indicating a stable anisotropic stellar structure.
\item
Exhibiting a robust linear correlation, the EoS underscores the
intrinsic relationship between $\rho$ and $p_{r}$. Similarly,
$p_{t}$ adheres to a linear model as well. These linear
relationships remain uniform from the center to the star's outer
layers. The observed uniformity of $\omega_{r}$ and $\omega_{t}$
validates their compliance with the requisite criteria for PS
viability, thereby denoting enhanced stability (Figure \textbf{8}).
\item
The condition $\Gamma > \frac{4}{3}$ is fulfilled (Figure
\textbf{9}), affirming the structural stability of the PS.
\item
Stability is achieved when the sum of the forces $F_{(\mathbb{Q},
\mathcal{T})}$, $F_{a}$, $F_{g}$ and $F_{h}$ equals zero (Figure
\textbf{10}). This equilibrium results in a robust configuration for
the PS.
\end{itemize}

Within the theoretical framework of $f(\mathbb{Q}, \mathcal{T})$
gravity, we have meticulously constructed a resilient and physically
congruent model of a charged PS. The results demonstrate that this
theory meets the necessary stability conditions and exhibits
congruence with empirical observational data. The incorporation of
charge notably affects the star's structure, altering its radius,
mass and overall stability. Our analysis highlights variations in
$\rho$, $p_{r}$ and $p_{t}$, as well as anisotropy. The charged
model remains both stable and physically feasible, despite these
variations. Furthermore, these findings are consistent with
contemporary studies in the literature \cite{112}.

Our study investigates charged anisotropic PS within the framework
of $f(\mathbb{Q},\mathcal{T})$ gravity, providing a novel
perspective on the structure and stability of PS. By extending GR
through non-metricity and the trace of the EMT, our work explores
the effects of charge and anisotropy on the pulsar's stability and
structural properties. We present a comprehensive analysis of
mass-radius relationships, energy conditions, and stability
criteria, emphasizing the critical role of charge in the formation
of compact stars. Deb et al. \cite{a} employed the Einstein-Maxwell
framework to analyze anisotropic strange stars using the MIT bag
model EoS. They explored quark stars, whereas our study investigates
pulsars within the Einstein-Maxwell gravity framework, whereas our
study provides insights from $f(\mathbb{Q},\mathcal{T})$ gravity,
offering a modified gravitational interpretation of pulsar behavior.
Deb et al. \cite{a1} also extended the analysis to
$f(\mathbb{R},\mathcal{T})$ gravity, where $\mathbb{R}$ is the Ricci
scalar, demonstrating that certain stars can exceed the
Chandrasekhar limit. Our primary focus is on the role of charge in
pulsar stability. The work in \cite{a2} examined strange stars like
LMC X-4 in $f(T,\mathcal{T})$ gravity, where $T$ is the torsion
scalar, a torsion-based modification of GR. Das et al. \cite{a3}
investigated tidal Love numbers in anisotropic compact stars within
GR. While both studies discuss anisotropy, our work uniquely
examines how charge-induced anisotropy influences stability.
Furthermore, the study in \cite{a4} focused on charged anisotropic
compact objects, making it directly relevant to our work. Although
their work is based on a different theoretical framework, we
highlight key similarities in how charge influences the formation
and stability of compact objects. Another important study by Yousaf
et al. \cite{a5} investigated the structure and stability of compact
objects under modified gravity, aligning with the central theme of
our research. Additionally, the study in \cite{a6} examined compact
stars within GR using color-flavor locked quark matter. Their study
emphasizes the color-flavor locked EoS used in quark star models,
whereas our work investigates pulsars with anisotropic matter and
charge under a modified gravity paradigm.

\section*{Appendix A: Casuality Condition}

\renewcommand{\theequation}{A\arabic{equation}}
\setcounter{equation}{0}
\begin{eqnarray}\nonumber
v_{r}^{2}&=&-\bigg[(B +1) \big(12 e^{3  \gamma_{0}r^2}r^4
\big(q_{0}^2 r^4+A \big) B ^2 (4 B +3)+4 e^{2 \gamma_{0} r^2}r^2 B
\big(q_{0}^2\big(2 \gamma_{0}
\\\nonumber
&\times&B ^2-3\big(8 B ^2+15 B+6\big)\big) r^4+A \big(\gamma_{0}^2
\big(\big(20 r^2+2\big) B ^2+3 \big(5 r^2+2\big) B\big) r^4
\\\nonumber
&+&3 \gamma_{0}\big(-4 r^2 B ^2+\big(2-3 r^2\big) B +1\big) r^2+a
\big(\gamma_{0} \big(\big(2-4 r^2\big) B ^2- \big(r^2-2\big) B
\\\nonumber
&+&3\big) r^2+\big(2 B ^2+6 B +3\big)\big) r^2-(4 B +3) \big(\big(3
r^2+7\big) B+6\big)\big)\big)-A
\\\nonumber
&\times&\big(8 B ^2+15 B+6\big) \big((7 B+6)+2 a^2 r^4 B
+\gamma_{0}^2 r^4 \big(2 a r^2 B -3(3 B +4)\big)
\\\nonumber
&+&\gamma_{0} r^2 \big(-2 a^2 B r^4-3 a B r^2+ (7 B
+6)\big)\big)+e^{\gamma_{0} r^2}\big(q_{0}^2 r^4 \big(56 B ^3+153 B
^2
\\\nonumber
&+&132 B +36\big)+A \big(-8 a (a-\gamma_{0}) \gamma_{0} B ^2 (4 B
+3) r^8-4 \gamma_{0}  B  (4 B +3) (a B
\\\nonumber
&+&3 \gamma_{0} (3 B +4)) r^6-12 (a-5 \gamma_{0})\eta \big(2 B ^2+3
B +1\big) r^4+(7 B +6) \big(8 \big(2 r^2
\\\nonumber
&+&1\big) B ^2+3 \big(4 r^2+5\big) B
+6\big)\big)\big)\big)\bigg]\bigg[4 e^{3 \gamma_{0} r^2}r^4 B ^2 (4
\eta +3) \big(q_{0}^2 r^4 (7 B +3)-A
\\\nonumber
&-&(B +3)\big)-A-\big(8 B ^2+15 B +6\big) \big(2 a^2 \big(\gamma_{0}
r^2-\big) B (3 B +2) r^4-a \gamma_{0}
\\\nonumber
&-&\big(2 \gamma_{0} B (3 B +2) r^2+3 \big(5 B ^2+13 B +6\big)\big)
r^4- \big(\gamma_{0}^2 \big(13 B ^2+21 B +6\big) r^4
\\\nonumber
&+&\gamma_{0}\big(13 B ^2+17 B +6\big) r^2+\big(13 B ^2+17 B
+6\big)\big)\big)+e^{\gamma_{0} r^2}\big(q_{0}^2 r^4\big(24 B ^4
\\\nonumber
&+&149 B ^3+261 B ^2+168 B +36\big)-A \big(-8 a^2 B (4 B +3)
\big(\gamma_{0} B (3 B +2) r^2
\\\nonumber
&+&R^2 \big(8 B ^2+15 B +6\big)\big) r^6+4 a \big(2 \gamma_{0}^2 B
^2 \big(12 B ^2+17 B +6\big) r^4+\gamma_{0}B \big(148 B ^3
\\\nonumber
&+&403 B ^2+339 B+90\big) r^2-3(B +1)^2 \big(22 B ^2+23 B
+6\big)\big) r^4+\big(4 \gamma_{0}^2 B
\\\nonumber
&\times& \big(52 B ^3+123 B ^2+87 B +18\big) r^6 +4 \gamma_{0}
\big(74 B ^4+141 B ^3+54 B ^2-33 B
\\\nonumber
&-&18\big) r^4+\big(13 B ^2+17 B +6\big) \big(8 \big(2 r^2+1\big) B
^2+3 \big(4 r^2 +5\big) B +6\big)\big)\big)\big)
\\\nonumber
&\times&e^{2 \gamma_{0} r^2}r^2 \big(-2 a^2 \big(\gamma_{0}
r^2-\big) A B ^2 (4 B +3)^2 r^6+a A \big(2 \gamma_{0}^2 B ^2 (4 B
+3)^2 r^6
\\\nonumber
&+&\gamma_{0}\big(\big(22-140 r^2\big) B ^4-5 \big(49 r^2-20\big) B
^3-3 \big(43 r^2-49\big) B ^2+\big(87
\\\nonumber
&-&18 r^2\big) B+18\big) r^2+\big(22 B ^4+ B ^3+147 B ^2+87 B
+18\big)\big) r^2+\big(\gamma_{0}^2 A
\\\nonumber
&+&\big(\big(22-4 r^2\big) B ^4+\big(100-23 r^2\big) B
^3+\big(147-39 r^2\big) B ^2 +\big(87-18 r^2\big) B
\\\nonumber
&+&18\big) r^4+\gamma_{0}\big(2 q_{0}^2 B ^2\big(11 B ^2+17 \eta
+6\big) r^4 +A \big(-4 r^2 B ^4+\big(-15 r^2\big) B ^3
\\\nonumber
&-&9 \big(r^2-15\big) B ^2+87 B+18\big)\big) r^2B \big(q_{0}^2 r^4
\big(56 B ^3+ B ^2+ B +18\big)-A
\\\label{38}
&\times&(4 B +3) \big(\big(r^2+13\big) B ^2+\big(3 r^2+17\big) B
+6\big)\big)\big)\big)\bigg]^{-1},
\end{eqnarray}
\begin{eqnarray}\nonumber
v_{t}^{2}&=&\bigg[(B +1) \big(12 e^{3 \gamma_{0} r^2} r^4
\big(q_{0}^2 r^4+A \big) B ^2 (4 B +3)+4 e^{2 \gamma_{0} r^2}r^2 B
\big(q_{0}^2\big(2 \gamma_{0} r^2 B ^2
\\\nonumber
&-&3\big(8 B ^2+15 B +6\big)\big) r^4+A  \big(\gamma_{0}^2
\big(\big(20 r^2+2\big) B ^2+3 \big(5 r^2+2\big) B +3\big) r^4
\\\nonumber
&+&3 \gamma_{0}\big(-4 r^2 B ^2+\big(2-3 r^2\big) B +1\big) r^2+a
\big(\gamma_{0} \big(\big(2-4 r^2\big) B ^2-3 \big(r^2-2\big) B
\\\nonumber
&+&3\big) r^2+\big(2 B ^2+6 B +3\big)\big) r^2-(4 B +3) \big(\big(3
r^2+7\big) B +\big)\big)\big)-A \big( B ^2
\\\nonumber
&+&15 B +6\big) \big((7 B +6)+2 a^2 r^4 B+\gamma_{0}^2 r^4 \big(2 a
r^2 B-3(3 B +4)\big)+\gamma_{0} r^2
\\\nonumber
&\times&\big(-2 a^2 B r^4-3 a B r^2+(7 B +6)\big)\big)+e^{\gamma_{0}
r^2}\big(q_{0}^2 r^4 \big(56 B ^3+ B ^2 +132 B
\\\nonumber
&+&36\big)+A \big(-8 a (a-\gamma_{0}) \gamma_{0} B ^2 (4 B +3) r^8-4
\gamma_{0}B (4 B +3) (a B +3 \gamma_{0}
\\\nonumber
&\times&(3 B +4))r^6-12 (a-5 \gamma_{0})B \big(2 B ^2+3 B +1\big)
r^4+(7 B +6) \big(8
\\\nonumber
&\times&\big(2 r^2+1\big)B ^2+3 \big(4 r^2+5\big) B
+6\big)\big)\big)\big)\bigg]\bigg[4 e^{3 \gamma_{0} r^2} r^4 B ^2 (4
B +3) \big(q_{0}^2 (5 B +3) r^4
\\\nonumber
&+&A B \big)+A \big(8 B ^2+15 B +6\big) \big(a^2 \big(\gamma_{0}
r^2-\big) \big(6 B ^2+13 B +6\big) r^4+a \gamma_{0}
\\\nonumber
&\times&\big(3\big(7 B ^2+14 B+6\big)-r^2 \big(6 B ^2+13
B+6\big)\big) r^4-\big(\gamma_{0}^2 \big(B ^2+6 B +6\big) r^4
\\\nonumber
&+&\gamma_{0}B (B +2) r^2+B (B +2)\big)\big) +e^{\gamma_{0}
r^2}\big(q_{0}^2 r^4 \big(72 B ^4+271 B ^3+357 B ^2
\\\nonumber
&+&192 B+36\big) R^6+A  B \big(-4 a^2 (4 B +3) \big(\gamma_{0} r^2
\big(6 B ^2+13 B +6\big)-\big(8 B ^2
\\\nonumber
&+&15 B +6\big)\big) r^6+4 a \big(\gamma_{0}^2 \big(24 B ^3+70 B
^2+63 B +18\big) r^4-\gamma_{0}\big(92 B ^3+245 B ^2
\\\nonumber
&+&204 B +54\big) r^2+3(B +1)^2 (2 B +1)\big) r^4+\big(4
\gamma_{0}^2 \big(4 B ^3+27 B ^2+42 B
\\\nonumber
&+&18\big) r^6+4 f\big(2 B ^3+9 B ^2+9 B +3\big) r^4+(B +2) \big(8
\big(2 r^2+1\big) B ^2+3
\\\nonumber
&+&\big(4 r^2+5\big) B +6\big)\big)\big)\big)-4 e^{2 \gamma_{0}
r^2}r^2 B \big(q_{0}^2 r^4 \big(\big(40 B ^3+99 B ^2+75 B+18\big)
\\\nonumber
&+&2 \gamma_{0} r^2 B ^2 (B +1)\big)+A \big(-a^2 \big(\gamma_{0}
r^2-R^2\big) B (4 B +3)^2 r^6+a \big(\gamma_{0}^2 B
\\\nonumber
&+&(4 B +3)^2+r^6+\gamma_{0}\big(\big(2-52 r^2\big) B ^3+\big(8-79
r^2\big) B ^2+\big(9-30 r^2\big) B
\\\nonumber
&+&3\big) r^2+\big(2 B ^3+8 B ^2+9 B +3\big)\big) r^2+
\big(\gamma_{0}^2 \big(\big(4 r^2+2\big) B ^3 +\big( r^2+8\big) B ^2
\\\nonumber
&+&\big(6 r^2+9\big) B +3\big) r^4+\gamma_{0} \big(4 r^2 B ^3+3
\big(r^2+2\big) B ^2+9 B +3\big) r^2+B (4 B
\\\label{39}
&+&3) \big(B  r^2+B +2\big)\big)\big)\big)\bigg]^{-1}.
\end{eqnarray}

\section*{Appendix B: Adiabatic Index}
\renewcommand{\theequation}{B\arabic{equation}}
\setcounter{equation}{0}
\begin{eqnarray}\nonumber
\Gamma&=&\frac{4}{3} \big(\bigg[1-3\big(8 \big(e^{\gamma_{0}
r^2}r^2-1\big) B ^2+3 \big(2 e^{\gamma_{0} r^2}r^2-5\big) B 6\big)
\big(2 e^{2 \gamma_{0} r^2}r^2
\\\nonumber
&\times&\big(2 \gamma_{2} A \big(2 B ^2+3 B +1\big) r^2+2 \gamma_{0}
A \big(2 B ^2+3 B +1\big) r^2 +R^2 B  \big(2 q_{0}^2 (2 B +1) r^4
\\\nonumber
&+&A \big)\big)+A \big(\gamma_{2}^2 (3 B +2) r^4+\gamma_{2} \big(2
R^2 \big(6 B ^2+12 B +5\big)-\gamma_{0} r^2 (3 B +2)\big) r^2
\\\nonumber
&+&\big(\gamma_{0} B (4 B +5) r^2+\big(4 \eta ^2+5 B
+2\big)\big)\big)+e^{\frac{\gamma_{0} r^2}{R^2}} \big(-2 q_{0}^2 r^4
\big (2 B ^2+5 B\big)
\\\nonumber
&-&A \big(2 \gamma_{2}^2 \eta (4 B +3) r^6+2 \gamma_{2} \big( \big(2
\big(6 \eta ^2+6 B +1\big) r^2+2 \eta ^2+3 B +1\big)-\gamma_{0} r^4
\eta
\\\nonumber
&\times&(4 B +3)\big) r^2+\big(2 \gamma_{0} \big((B +2) r^2+2 B ^2+
\eta +1\big) r^2+\big(B ^2+\big( r^2+5\big)
\\\nonumber
&\times&B +2\big)\big)\big)\big)\big)\bigg]\bigg[2 \big(4 e^{3
\gamma_{0} r^2}r^4 \eta ^2 (4 B +3) \big(q_{0}^2 r^4 (7 B +3)-A
(\eta +3)\big)-A8 B ^2
\\\nonumber
&\times&\big(+15 B+6\big) \big(2 \gamma_{2}^2 \big(\gamma_{0}
r^2-R^2\big) \eta (3 B +2) r^4-\gamma_{2} \gamma_{0} \big(2
\gamma_{0} B (3 \eta +2) r^2+3
\\\nonumber
&\times&\big(+6\big)\big) r^4-\big(\gamma_{0}^2 \big(13 \eta ^2+21 B
+6\big) r^4+\gamma_{2}\big(13 \eta ^2+17 B +6\big) r^2+
\\\nonumber
&\times&\big(13 \eta ^2+6\big)\big)\big)+e^{\gamma_{0}
r^2}\big(q_{0}^2 r^4 R^6 \big(24 B ^4+149 \eta ^3+261 B ^2+168 B
+36\big)-A \big(
\\\nonumber
&+&R^2 \big(8 \eta ^2+15 B +6\big)\big) r^6+4 a \big(2 \gamma_{0}^2
B ^2 \big(12 B ^2+17 B +6\big) r^4+\gamma_{0}B \big (148 B ^3
\\\nonumber
&-&3(B +1)^2 \big(22 B ^2+23 B +6\big)\big) r^4+\big(4 ^\gamma_{0}2
B \big(52 B ^3+123 B ^2+87 B +18\big) r^6
\\\nonumber
&+&4 f R^2 \big(74 \eta ^4+141 B ^3+54 B ^2-33 B\big) r^4 \big(8
\big(2 r^2+1\big) \eta ^2+3 \big(4 r^2+5\big) B
+6\big)\big)\big)\big)
\\\nonumber
&-&4 e^{2 \gamma_{0} r^2}r^2 \big(-2 \gamma_{2}^2 \big(\gamma_{0}
r^2-R^2\big) A B ^2 (4 \eta
 +3)^2 r^6 +\gamma_{2} A \big(2 \gamma_{0}^2 B
^2 (4 B +3)^2 r^6
\\\nonumber
&+& \gamma_{0}\big(\big(22-140 r^2\big)B
 ^4 -5 \big(49
r^2-20\big) B ^3-3 \big(43 r^2-49\big)B ^2+\big(87-18 r^2\big)
\\\nonumber &+&18\big) r^2+\big(22B
^4+100B ^3+147 B ^2+87 B +18\big)\big) r^2+\big(\gamma_{0}^2 A
\big(\big(22-4 r^2\big) B ^4
\\\nonumber
&+&\big(147-39 r^2\big) B ^2+\big(87-18 r^2\big) B +18\big)
r^4+\gamma_{0}\big(2 q_{0}^2 B ^2 \big(11 B ^2+17 B +6\big) r^4
\\\nonumber
&+&\big(66-15 r^2\big) B ^3-9 \big(r^2-15\big) B ^2+87 B
+18\big)\big) r^2+B \big(j^2 r^4 \big(56 B ^3+129 B ^2
\\\nonumber
&-&A (4 B +3) \big(\big(r^2+13\big) B ^2+\big(3 r^2+17\big) B
+6\big)\big)\big)\big)\big)\bigg]^{-1}
\end{eqnarray}
\begin{eqnarray}\nonumber
\Gamma_{r}&=&-\bigg[2 (B +1) \big(B  \big(2 e^{\gamma_{0} r^2} (4 B
+3) r^2-8 B -15\big)-6\big) \big(-\gamma_{2}^2 \big(2 e^{\gamma_{0}
r^2} r^2+1\big) A  B
\\\nonumber
&\times&(4 B +3) r^4+\gamma_{2} A \big(\gamma_{0} \big(2
e^{\gamma_{0} r^2}r^2+1\big) B (4 B +3) r^2\big(6 e^{2 \gamma_{0}
r^2}\big( B ^2+ B \big)
\\\nonumber
&\times& r^2+10 B ^2+21 B -e^{\gamma_{0} r^2}\big(\big(28 \eta ^2+30
B +6\big) r^2+6 B ^2+9 B\big)+9\big)\big) r^2
\\\nonumber
&+&R^2 \big(\gamma_{0} A  \big(6 e^{2 \gamma_{0} r^2}\big(2 B ^2+3 B
+1\big) r^2+2 B ^2+e^{\gamma_{0} r^2}\big(r^2 \big(4 B ^2-6\big)-3
\big(2 \eta ^2
\\\nonumber
&+&3 B +1\big)\big)-3\big) r^2+\big(2 e^{\frac{\gamma_{0} r^2}{R^2}}
q_{0}^2 \big(2 e^{\gamma_{0} r^2}r^2+1\big) \eta ^2
r^4+\big(-1+e^{\gamma_{0} r^2}\big)
\\\nonumber
&\times& A \big(2 \big(2 e^{\gamma_{0} r^2} r^2-5\big) B ^2+3 \big(2
e^{\gamma_{0} r^2}r^2-5\big) B -6\big)\big)\big)\big)\big(12 e^{3
\gamma_{0}r^2}r^4 \big(q_{0}^2 r^4
\\\nonumber
&+&A \big) \eta ^2 (4 B +3)+4 e^{2 \gamma_{0} r^2}r^2 B\big(q_{0}^2
\big(2 \gamma_{0} r^2B^2-3\big(B ^2+B+6\big)\big) r^4
\\\nonumber
&+&A \big(\gamma_{0}^2 \big(\big(20 r^2+2\big) B ^2+3 \big(5
r^2+2\big) \eta 3\big) r^4+3 \gamma_{0}\big(-4 r^2 B ^2+\big(2-3
+r^2\big)
\\\nonumber
&\times&B +1\big)-r^2+\gamma_{2} \big(\gamma_{0} \big(\big(4
r^2\big) B ^2-3 \big(r^2-2\big) \eta +3\big) r^2+\big(2 B ^2+B
+3\big)\big)
\\\nonumber
&\times&r^2-(4 B +3) \big (\big(3 r^2+7\big) B +6\big)\big)\big)
R^2-A \big(8 \eta ^2+15 B +6\big) \big((7 B +6)
\\\nonumber
&+&2 a^2 r^4 \eta+\gamma_{0}^2 r^4 \big(2 a r^2 B -3 3 B
+4)\big)+\gamma_{0} r^2 \big(-2 \gamma_{2}^2 \eta r^4-3 \gamma_{2}
R^2 B r^2
\\\nonumber
&+&(7 B +6)\big)\big)+e^{\gamma_{0} r^2} \big(q_{0}^2 r^4 \big(56 B
^3+153 B ^2+132 B +36\big)+A \big(-8 \gamma_{2} (\gamma_{2}
\\\nonumber
&-&\gamma_{0}) \gamma_{0} B ^2 (4 B +3) r^8-4 \gamma_{0}B (4 B +3)
(a B +3 \gamma_{0} (3 \eta +4)) r^6-12 (\gamma_{2}-5 )
\\\nonumber
&-& \big(2 B ^2+3 B +1\big) r^4+(7 B +6) \big(8 \big(2 r^2+1\big)
\eta ^2+3 \big(4 r^2+5\big) B +6\big)\big)\big)\big)\bigg]
\\\nonumber
&\times&\bigg[\big(8 \big(e^{\gamma_{0} r^2}r^2-1\big) B ^2+3 \big(2
e^{\gamma_{0} r^2}r^2-5\big) B -6\big) \big(4 e^{3 \gamma_{0}
r^2}r^4 B ^2 (4 B +3) \big(q_{0}^2 r^4
\\\nonumber
&\times&(7 B +3)-A (\eta +3)\big)-A \big(8 B ^2+15 B +6\big) \big(2
\gamma_{2}^2 \big(\gamma_{0} r^2-\big) \eta (3 \eta +2) r^4
\\\nonumber
&-&a \gamma_{0} \big(2 \gamma_{0} B (3 B +2) r^2+3 \big(5 \eta ^2+13
B +6\big)\big) r^4- \big(\gamma_{0}^2 \big(13 \eta ^2+21 B +6\big)
r^4
\\\nonumber
&+&\gamma_{0} R^2 \big (13 \eta ^2+17 B+6\big) r^2+\big(13 B ^2+17
\eta +6\big)\big)\big)+e^{\gamma_{0} r^2}\big(q_{0}^2 r^4\big(24
\eta ^4
\\\nonumber
&+&149 B ^3+261 B ^2+168 B+36\big)-A \big(-8 \gamma_{2}^2 \eta (4 B
+3) \big(\gamma_{0} B (3B +2) r^2
\\\nonumber
&+&\big(8 B ^2+15 \eta +6\big)\big) r^6+4 \gamma_{2} \big(2
\gamma_{0}^2 B ^2 \big(12 \eta ^2 +17 B +6\big) r^4+\gamma_{0}B
\big(B^3
\\\nonumber
&+&403 B ^2+339 B +90\big) r^2-3(\eta +1)^2 \big(22 B ^2+23 \eta
+6\big)\big) r^4+\big(4 \gamma_{0}^2 B
\\\nonumber
&\times&\big(52 B ^3+123 B ^2+87 B +18\big) r^6+4 \gamma_{0} \big(74
B ^4+141 B ^3+54 B ^2-33 B
\\\nonumber
&-&18\big) r^4+\big(13 B ^2+17 B+6\big) \big(8 \big(2 r^2+1\big)
\eta ^2+3 \big(4 r^2+5\big) B +6\big)\big)\big)\big)
\\\nonumber
&-&4 e^{2 \gamma_{0} r^2}r^2 \big(-2 \gamma_{2}^2 \big(\gamma_{0}
r^2-\big) A B ^2 (4 B +3)^2 r^6+a A \big(2 \gamma_{0}^2 B ^2 (4 B
+3)^2 r^6+\gamma_{0}
\\\nonumber
&-&5 \big(49 r^2-20\big) B ^3-3 \big(43 r^2-49\big) B ^2 \big(87-18
r^2\big) \eta +18\big) r^2+\big(22 B ^4
\\\nonumber
&+&100 B ^3+147 \eta ^2+87 B +18\big)\big) r^2+\big(^\gamma_{0}2 A
\big(\big(22-4 r^2\big) \eta ^4+\big(100-23 r^2\big) B ^3
\\\nonumber
&+&\big(147-39 r^2\big) \eta ^2+\big(87-18 r^2\big) B +18\big)
r^4+\gamma_{0}\big(2 q_{0}^2 B ^2 \big(11 B ^2+17 \eta +6\big) r^4
\\\nonumber
&+&A \big(-4 r^2 B ^4+\big(66-15 r^2\big) \eta ^3-9 \big(r^2-15\big)
B ^2+87 \eta +18\big)\big) r^2+B \big(q_{0}^2 r^4
\\\nonumber
&\times&\big(56 B^3+129 B ^2+87 B +18\big)-A (4 B +3)
\big(\big(r^2+13\big) B ^2+\big(3
r^2+17\big)\big)\big)\big)\big)\big)
\\\nonumber
&\times& \big(2 e^{\frac{2 \gamma_{0} r^2}{R^2}} r^2 \big(\gamma_{2}
A (B +1)r^2+\gamma_{0} A (B +1) (11 B +6) r^2+R^2 B \big(q_{0}^2
(7B+3) r^4
\\\nonumber
&+&A (B +3)\big)\big)+A \big(-2 \gamma_{2}^2 \eta (3 B +2) r^4
+\gamma_{2} \big(2 \gamma_{0} \eta (3 B +2) r^2+(B (21B\\\nonumber
&+&43)+18)\big) r^2+\big(\gamma_{0} (B (13 \eta +21)+6) r^2+ (B (13
B +17)+6)\big)\big )
\\\nonumber
&-&e^{\gamma_{0} r^2}\big(q_{0}^2 r^4 (B(3B+13)+6)+A \big(4
\gamma_{2}^2 B  (4 B +3) r^6+\gamma_{2} \big(\big(2 (B (27 B
\\\nonumber
&+&29)+6) r^2+(B +1) (11 B +6)\big)-4 \gamma_{0} r^4 B (4 B +3)\big)
r^2+\big(\gamma_{0} \big(2 (B+2)
\\\nonumber
&+&(B +1) (11B+6)\big) r^2+\big(B \big(2 (B+3) r^2+13 B
+17\big)+6\big)\big)\big)\big)\big)\bigg]^{-1},
\end{eqnarray}
\begin{eqnarray}\nonumber
\Gamma_{t}&=&-\bigg[2 (B +1) \big(\gamma_{0}^2 \big(2 \gamma_{2}
\eta \big(8 \big(2 e^{\gamma_{2} r^2}r^2-1\big) B ^2+3 \big(4
e^{\gamma_{0} r^2}r^2-5\big)B -6\big) r^2
\\\nonumber
&+&R^2 \big(8 \big(-18 e^{\gamma_{0} r^2}r^2+e^{2 \gamma_{0}
r^2}\big(10 r^4+r^2\big)+9\big) B ^3+3 \big(-100 e^{\gamma_{0}
r^2}r^2
\\\nonumber
&+&4 e^{\frac{2 \gamma_{0} r^2}{R^2}} \big(5 r^2+2\big)
r^2+77\big)B^2+6 \big(-24 e^{\gamma_{0} r^2}r^2+2 e^{2
\gamma_{0}r^2}r^2+39\big) \eta +72\big)\big) r^4
\\\nonumber
&+&\gamma_{0} \big(-2 a^2 B \big(8 \big(2 e^{\gamma_{0} r^2}
r^2-1\big) \eta ^2+3 \big(4 e^{\gamma_{0} r^2} r^2-5\big) B -6\big)
r^4-a\eta \big(4 e^{\gamma_{0} r^2}
\\\nonumber
&\times& B  (4 \eta +3) r^2+4 e^{2 \gamma_{0} r^2} \big(\big(4
r^2-2\big) B ^2+3 \big(r^2-2\big) B -3\big) r^2-3 \big(8 B ^2
\\\nonumber
&+&15 B +6\big)\big) r^2+\big(-8 \big(6 e^{2 \gamma_{0}r^2} r^4-15
e^{\gamma_{0} r^2}r^2+7\big) B ^3-3 \big(-60 e^{\gamma_{0} r^2}r^2
\\\nonumber
&+&4 e^{2 \gamma_{0} r^2}\big(3 r^2-2\big) r^2+51\big) B ^2+12
\big(5 e^{\gamma_{0} r^2}r^2+e^{2 \gamma_{0} r^2} r^2-11\big) B
-36\big)\big) r^2
\\\nonumber
&+&R^2 \big(-2 \gamma_{2}^2 B \big(8 B ^2+15 B +6\big) r^4+4
\gamma_{2} e^{\gamma_{0} r^2}\eta \big(e^{\gamma_{0} r^2} \big(2 B
^2+6 B +3\big)
\\\nonumber
&-&3 \big(2B^2+3B+1\big)\big) r^4+\big(-1+e^{\gamma_{0} r^2}\big)
R^4 \big(8 \big(6 e^{2 \gamma_{0} r^2}r^4-14 e^{\gamma_{0}
r^2}r^2+7\big) B ^3
\\\nonumber
&+&9 \big(4 e^{2 \gamma_{0} r^2}r^4-20 e^{\gamma_{0} r^2}r^2+17\big)
B ^2+\big(132-72 e^{\gamma_{0} r^2}r^2\big) B +36\big)\big)\big)
\big(-a^2
\\\nonumber
&\times& \big(2 e^{\gamma_{0} r^2}r^2+1\big) B  (4 B +3) r^4+a
\big(\gamma_{0} \big(2 e^{\gamma_{0} r^2} r^2+1\big) B (4 B +3)
r^2+\big(6 e^{2 f r^2}
\\\nonumber
&\times& \big(2 B ^2+3 B +1\big) r^2+10 B ^2+21 B -e^{\gamma_{0}
r^2}\big(\big(28 B ^2+30 B +6\big) r^2+6 B ^2
\\\nonumber
&+&9 B +3\big)+9\big)\big) r^2+\big(\gamma_{0} \big(6 e^{2
\gamma_{0} r^2}\big(2 \eta ^2+3 B +1\big) r^2+2 B ^2+e^{\gamma_{0}
r^2}\big(r^2
\\\nonumber
&\times& \big(4B ^2-6\big)-3 \big(2 B ^2+3 B +1\big)\big)-3\big)
r^2+\big(-1+e^{\gamma_{0} r^2}\big) R^2 \big(2 \big(2 e^{\gamma_{0}
r^2}r^2
\\\nonumber
&-&5\big)B^2+3 \big(2 e^{\gamma_{0} r^2} r^2-5\big) B
-6\big)\big)\big)\bigg]\bigg[\big(a^2 \big( \big(16 \big(4 e^{2
\gamma_{0} r^2}r^4-8 e^{\gamma_{0} r^2}r^2+3\big)B^4
\\\nonumber
&+&2 \big(48 e^{2 \gamma_{0} r^2}r^4-168 e^{\gamma_{0} r^2}
r^2+97\big) B ^3+3 \big(12 e^{2 \gamma_{0} r^2}r^4-92 e^{\gamma_{0}
r^2} r^2+93\big)B^2
\\\nonumber
&-&24 \big(3 e^{\gamma_{0} r^2}r^2-7\big) B +36\big)-\gamma_{0} r^2
\big(16 \big(4 e^{2 \gamma_{0} r^2}r^4-6 e^{\gamma_{0}
r^2}r^2+3\big)B^4+2 \big(48
\\\nonumber
&\times& e^{2 \gamma_{0} r^2}r^4-140 e^{\gamma_{0} r^2} r^2+97\big)
\eta ^3+9 \big(4 e^{2 \gamma_{0} r^2} r^4-28 e^{\gamma_{0}
r^2}r^2+31\big) B ^2
\\\nonumber
&-&24 \big(3 e^{\gamma_{0} r^2}r^2-7\big) \eta +36\big)\big) r^4+a
\big(4 e^{\gamma_{0} r^2} B (\eta +1) \big(e^{\gamma_{0} r^2}\big(2
B ^2+6 \eta +3\big)
\\\nonumber
&-&3 \big(2 B ^2+3 B +1\big)\big)-\gamma_{0} \big(8 \big(-46
e^{\gamma_{0} r^2}r^2+e^{ 2 \gamma_{0} r^2}\big(26 r^2-1\big)
r^2+21\big) B^4
\\\nonumber
&\times&\big(-980 e^{\gamma_{0} r^2}r^2+4 e^{2 \gamma_{0}
r^2}\big(79 r^2-8\big) r^2+651\big) B ^3+12 \big(-68 e^{\gamma_{0}
r^2} r^2
\\\nonumber
&+&e^{2 \gamma_{0} r^2}\big(10 r^2-3\big) r^2+75\big)B^2-6 \big(36
e^{\gamma_{0} r^2}r^2+2 e^{2 \gamma_{0} r^2}r^2-87\big) \eta
+108\big) R^2
\\\nonumber
&+&f^2 r^2 \big(16 \big(4 e^{2 \gamma_{0}r^2}r^4-6 e^{\gamma_{0}
r^2}r^2+3\big)B^4+2 \big(48 e^{2 \gamma_{0} r^2}r^4-140
e^{\gamma_{0} r^2}r^2+97\big) B ^3
\\\nonumber
&+&9 \big(4 e^{2 \gamma_{0} r^2}r^4-28 e^{\gamma_{0}
r^2}r^2+31\big)B^2-24 \big(3 e^{\gamma_{0} r^2} r^2-7\big) \eta
+36\big)\big) r^4+\big(\gamma_{0}^2
\\\nonumber
&\times& \big(8 \big(-2 e^{\gamma_{0} r^2}r^2+e^{2 \gamma_{0} r^2}
\big(2 r^4+r^2\big)+1\big) B ^4+\big(-108 e^{A_{1} r^2}r^2+4 e^{2
\gamma_{0} r^2}
\\\nonumber
&\times& \big(11 r^2+8\big) r^2+63\big) \eta ^3+12 \big(-14
e^{\gamma_{0} r^2}r^2+e^{2 \gamma_{0} r^2}\big(2 r^2+3\big)
r^2+12\big) \eta ^2
\\\nonumber
&+&6 \big(-12 e^{\gamma_{0} r^2}r^2+2 e^{2 \gamma_{0}r^2}
r^2+21\big) \eta+36\big) r^4+\gamma_{0}B \big(8B^3+31 B ^2+36 \eta
\\\nonumber
&-&4 e^{\gamma_{0} r^2}r^2 \big(2 B^3+9 B ^2+9 \eta +3\big)+4 e^2
\gamma_{0} r^2r^2 \big(4 r^2 B ^3+3 \big(r^2+2\big) B ^2+9
\eta+3\big)
\\\nonumber
&+&12\big) r^2-\big(-1+e^{\gamma_{0} r^2}\big) R^4 B \big(8 \big(2
e^{2 r^2}r^4-2 e^{\gamma_{0} r^2} r^2+1\big) B^3+\big(12 e^{2
\gamma_{0} r^2} r^4
\\\nonumber
&-&44 e^{\gamma_{0} r^2}r^2+31\big) B ^2+\big(36-24 e^{\gamma_{0}
r^2}r^2\big) B +12\big)\big)\big) \big(-2 a^2 B \big(2 e^{\gamma_{0}
r^2}(4 B +3) r^2
\\\nonumber
&+&3 B +2\big) r^4+a \big(2 \gamma_{0} B \big(2 e^{\gamma_{0} r^2}
(4 B +3) r^2+3 B +2\big) r^2+\big(2 e^{2 f r^2} \big(11 B ^2
\\\nonumber
&+&17 B +6\big) r^2+21 \eta ^2+43 B -e^{\gamma_{0} r^2} \big(2
\big(27 B ^2+29 \eta +6\big) r^2+11 B ^2+17 \eta +6\big)
\\\nonumber
&+&18\big)\big) r^2+\big(\gamma_{0} \big(2 e^{2 \gamma_{0}
r^2}\big(11 B ^2+17 \eta
 +6\big) r^2+13 B ^2+21 B
-e^{\gamma_{0} r^2}\big(2 \big(\eta ^2
\\\nonumber
&+&5 B +6\big) r^2+11 \eta ^2+17 B +6\big)+6\big)
r^2+\big(-1+e^{\gamma_{0} r^2}\big)\big(\big(2 e^{\gamma_{0} r^2}
r^2
\\\nonumber
&-&13\big) \eta ^2+\big(6 e^{\gamma_{0} r^2}
r^2-17\big)B-6\big)\big)\big)\bigg].
\end{eqnarray}
\\
\textbf{Data Availability Statement:} No new data were generated or
analyzed in support of this research.

\end{document}